\documentclass{aastex63}
\usepackage{changepage}
\usepackage[graphicx]{realboxes}

\newcommand{\Caltech}{\affil{California Institute of Technology, 1200 E. California Blvd., MC 249-17, Pasadena, CA 91125, USA}}

\newcommand{\Tamkang}{\affil{Department of Physics, Tamkang University, No.151, Yingzhuan Rd., Tamsui Dist., New Taipei City 251301, Taiwan}}

\newcommand{\GEMINI}{\affil{Gemini Observatory/NSF’s NOIRLab, 950 N. Cherry Avenue, Tucson, AZ, 85719, USA
}}
\newcommand{\ASCL}{\affil{Astrophysics Source Code Librar
y, Michigan Technological University, 1400 Townsend Drive, Houghton, MI 49931}}

\newcommand{\OSU}{\affil{Department of Astronomy, The Ohio State University, 140 West 18th Avenue, Columbus, Ohio 43210, USA}}

\newcommand{\Alberta}{\affil{Department of Physics, University of Alberta, Edmonton, AB T6G 2E1, Canada}}

\newcommand{\ANU}{\affil{Research School of Astronomy and Astrophysics, Australian National University, Canberra, ACT 2611, Australia}}

\newcommand{\IPARCOS}{\affil{Instituto de F\'{\i}sica de Part\'{\i}culas y del Cosmos, Universidad Complutense de Madrid, E-28040 Madrid, Spain}}

\newcommand{\IPAC}{\affil{Caltech-IPAC, 1200 E. California Blvd. Pasadena, CA 91125, USA}}

\newcommand{\Carnegie}{\affil{Observatories of the Carnegie Institution for Science, 813 Santa Barbara Street, Pasadena, CA 91101, USA}}

\newcommand{\CCAPP}{\affil{Center for Cosmology and AstroParticle Physics, 191 West Woodruff Avenue, Columbus, OH 43210, USA}}

\newcommand{\CfA}{\affil{Harvard-Smithsonian Center for Astrophysics, 60 Garden Street, Cambridge, MA 02138, USA}}

\newcommand{\CITEVA}{\affil{Centro de Astronomía (CITEVA), Universidad de Antofagasta, Avenida Angamos 601, Antofagasta, Chile}}

\newcommand{\ESO}{\affil{European Southern Observatory, Karl-Schwarzschild Stra{\ss}e 2, D-85748 Garching bei M\"{u}nchen, Germany}}

\newcommand{\Heidelberg}{\affil{Astronomisches Rechen-Institut, Zentrum f\"{u}r Astronomie der Universit\"{a}t Heidelberg, M\"{o}nchhofstra\ss e 12-14, D-69120 Heidelberg, Germany}}

\newcommand{\ICRAR}{\affil{International Centre for Radio Astronomy Research, University of Western Australia, 35 Stirling Highway, Crawley, WA 6009, Australia}}

\newcommand{\IRAM}{\affil{Institut de Radioastronomie Millim\'{e}trique (IRAM), 300 Rue de la Piscine, F-38406 Saint Martin d'H\`{e}res, France}}

\newcommand{\IRAP}{\affil{CNRS, IRAP, 9 Av. du Colonel Roche, BP 44346, F-31028 Toulouse cedex 4, France}}

\newcommand{\UPS}{\affil{Universit\'{e} de Toulouse, UPS-OMP, IRAP, F-31028 Toulouse cedex 4, France}}

\newcommand{\ITA}{\affil{Universit\"{a}t Heidelberg, Zentrum f\"{u}r Astronomie, Institut f\"{u}r Theoretische Astrophysik, Albert-Ueberle-Str 2, D-69120 Heidelberg, Germany}}

\newcommand{\IWR}{\affil{Universit\"{a}t Heidelberg, Interdisziplin\"{a}res Zentrum f\"{u}r Wissenschaftliches Rechnen, Im Neuenheimer Feld 205, D-69120 Heidelberg, Germany}}

\newcommand{\JHU}{\affil{Department of Physics and Astronomy, The Johns Hopkins University, Baltimore, MD 21218, USA}}

\newcommand{\MPE}{\affil{Max-Planck-Institut f\"{u}r extraterrestrische Physik, Giessenbachstra{\ss}e 1, D-85748 Garching, Germany}}

\newcommand{\MPIA}{\affil{Max-Planck-Institut f\"{u}r Astronomie, K\"{o}nigstuhl 17, D-69117, Heidelberg, Germany}}

\newcommand{\NRAO}{\affil{National Radio Astronomy Observatory, 520 Edgemont Road, Charlottesville, VA 22903-2475, USA}}

\newcommand{\OAN}{\affil{Observatorio Astron\'{o}mico Nacional (IGN), C/Alfonso XII, 3, E-28014 Madrid, Spain}}

\newcommand{\UToledo}{\affil{University of Toledo, 2801 W. Bancroft St., Mail Stop 111, Toledo, OH, 43606}}

\newcommand{\UBonn}{\affil{Argelander-Institut f\"ur Astronomie, Universit\"at Bonn, Auf dem H\"ugel 71, 53121 Bonn, Germany}}

\newcommand{\UChile}{\affil{Departamento de Astronom\'{i}a, Universidad de Chile, Camino del Observatorio 1515, Las Condes, Santiago, Chile}}

\newcommand{\UCM}{\affil{Departamento de F\'{\i}sica de la Tierra y Astrof\'{\i}sica, Universidad Complutense de Madrid, E-28040 Madrid, Spain}}

\newcommand{\ULyon}{\affil{Univ Lyon, Univ Lyon 1, ENS de Lyon, CNRS, Centre de Recherche Astrophysique de Lyon UMR5574,\\ F-69230 Saint-Genis-Laval, France}}

\newcommand{\UWyoming}{\affil{Department of Physics and Astronomy, University of Wyoming, Laramie, WY 82071, USA}}

\newcommand{\LAM}{\affil{
Aix Marseille Univ, CNRS, CNES, LAM (Laboratoire d’Astrophysique de Marseille),  F-13388 Marseille,
France}}

\newcommand{\UHawaii}{\affil{Institute for Astronomy, University of Hawaii, 2680 Woodlawn Drive, Honolulu, HI 96822, USA}}

\newcommand{\UGent}{\affil{Sterrenkundig Observatorium, Universiteit Gent, Krijgslaan 281 S9, B-9000 Gent, Belgium}}

\newcommand{\STScI}{\affil{Space Telescope Science Institute, 3700 San Martin Drive, Baltimore, MD 21218, USA}}

\newcommand{\INAF}{\affil{INAF -- Osservatorio Astrofisico di Arcetri, Largo E. Fermi 5, I-50157, Firenze, Italy}}

\newcommand{\LERMA}{\affil{Observatoire de Paris, PSL Research University, CNRS, Sorbonne Universit\'es, 75014 Paris}}

\newcommand{\SAIMSU}{\affil{Sternberg Astronomical Institute, Lomonosov Moscow State University, Universitetsky pr. 13, 119234 Moscow, Russia}}

\newcommand{\ngal}{38}
\newcommand{\ngalphangs}{74}
\newcommand{\msun}{M$_{\odot}$}
\newcommand{\sfr}{M$_{\odot}$ yr$^{-1}$}

\received{January 7, 2021}
\accepted{August 23, 2021}

\submitjournal{ApJS}

\shorttitle{The PHANGS-HST Survey}
\shortauthors{J.C. Lee et al.}

\begin{document}

\title{The PHANGS-HST Survey: \\ Physics at High Angular resolution in Nearby GalaxieS with the Hubble Space Telescope}

\correspondingauthor{Janice C. Lee}
\email{janice.lee@noirlab.edu}

\author[0000-0002-2278-9407]{Janice C. Lee}
\GEMINI

\author{Bradley C. Whitmore}
\STScI

\author[0000-0002-8528-7340]{David A. Thilker}
\JHU

\author[0000-0003-1943-723X]{Sinan Deger}
\Caltech

\author{Kirsten L. Larson}
\Caltech
\STScI

\author{Leonardo Ubeda}
\STScI

\author[0000-0002-5259-2314]{Gagandeep S. Anand}
\UHawaii
\STScI

\author{M\'ed\'eric Boquien}
\CITEVA

\author[0000-0003-0085-4623]{Rupali Chandar}
\UToledo

\author[0000-0002-5782-9093]{Daniel A. Dale}
\UWyoming

\author[0000-0002-6155-7166]{Eric Emsellem}
\ESO
\ULyon

\author[0000-0002-2545-1700]{Adam K. Leroy}
\OSU

\author[0000-0002-5204-2259]{Erik~Rosolowsky}
\Alberta

\author[0000-0002-3933-7677]{Eva Schinnerer}
\MPIA

\author[0000-0002-2617-5517]{Judy Schmidt}
\ASCL

\author{James Lilly}
\UWyoming

\author[0000-0003-2261-5746]{Jordan Turner}
\UWyoming

\author{Schuyler Van Dyk}
\IPAC

\author{Richard L. White}
\STScI

\author[0000-0003-0410-4504]{Ashley~T.~Barnes}
\UBonn

\author[0000-0002-2545-5752]{Francesco Belfiore}
\INAF

\author[0000-0003-0166-9745]{Frank Bigiel}
\UBonn

\author[0000-0003-4218-3944]{Guillermo A. Blanc}
\Carnegie
\UChile

\author[0000-0001-5301-1326]{Yixian Cao}
\LAM

\author[0000-0002-5635-5180]{Melanie Chevance}
\Heidelberg

\author[0000-0002-8549-4083]{Enrico Congiu}
\UChile

\author[0000-0002-4755-118X]{Oleg~V.~Egorov}
\Heidelberg
\SAIMSU

\author[0000-0001-6708-1317]{Simon C.~O. Glover}
\ITA

\author[0000-0002-3247-5321]{Kathryn Grasha}
\ANU

\author[0000-0002-9768-0246]{Brent Groves}
\ICRAR
\ANU

\author[0000-0001-9656-7682]{Jonathan D. Henshaw}
\MPIA

\author{Annie Hughes}
\IRAP
\UPS

\author[0000-0002-0560-3172]{Ralf S.\ Klessen}
\ITA
\IWR

\author[0000-0001-9605-780X]{Eric Koch}
\CfA
\Alberta

\author[0000-0001-6551-3091]{Kathryn Kreckel}
\Heidelberg

\author[0000-0002-8804-0212]{J.~M.~Diederik Kruijssen}
\Heidelberg

\author[0000-0001-9773-7479]{Daizhong Liu}
\MPIA

\author[0000-0002-1790-3148]{Laura A. Lopez}
\OSU \CCAPP

\author[0000-0002-5993-6685]{Ness Mayker}
\OSU 
\CCAPP

\author[0000-0002-6118-4048]{Sharon E. Meidt}
\UGent

\author[0000-0001-7089-7325]{Eric J. Murphy}
\NRAO

\author[0000-0002-1370-6964]{Hsi-An Pan}
\MPIA
\Tamkang

\author[0000-0003-3061-6546]{Jérôme Pety}
\IRAM
\LERMA

\author[0000-0002-0472-1011]{Miguel Querejeta}
\OAN

\author[0000-0001-7876-1713]{Alessandro Razza}
\UChile

\author[0000-0002-2501-9328]{Toshiki Saito}
\MPIA

\author[0000-0003-0651-0098]{Patricia S\'anchez-Bl\'azquez}
\UCM
\IPARCOS

\author[0000-0002-6363-9851]{Francesco Santoro}
\MPIA

\author[0000-0002-5783-145X]{Amy Sardone}
\OSU \CCAPP

\author{Fabian Scheuermann}
\Heidelberg

\author{Andreas Schruba}
\MPE

\author[0000-0003-0378-4667]{Jiayi~Sun}
\OSU

\author[0000-0003-1242-505X]{Antonio Usero}
\OAN

\author{E. Watkins}
\Heidelberg

\author[0000-0002-0012-2142]{Thomas G. Williams}
\MPIA

\begin{abstract}
The PHANGS program is building the first dataset to enable the multi-phase, multi-scale study of star formation across the nearby spiral galaxy population.  This effort is enabled by large survey programs with ALMA, VLT/MUSE, and HST, with which we have obtained CO(2--1) imaging, optical spectroscopic mapping, and high resolution UV-optical imaging, respectively.  Here, we present PHANGS-HST, which has obtained five band NUV-U-B-V-I imaging of the disks of \ngal\ spiral galaxies at distances of 4--23 Mpc, and parallel V and I band imaging of their halos, to provide a census of tens of thousands of compact star clusters, and multi-scale stellar associations.  The combination of HST, ALMA, and VLT/MUSE observations will yield an unprecedented joint catalog of the observed and physical properties of $\sim$100,000 star clusters, associations, HII regions, and molecular clouds.  With these basic units of star formation, PHANGS will systematically chart the evolutionary cycling between gas and stars, across a diversity of galactic environments found in nearby galaxies.  We discuss the design of the PHANGS-HST survey, and provide an overview of the HST data processing pipeline and first results, highlighting new methods for selecting star cluster candidates, morphological classification of candidates with convolutional neural networks, and identification of stellar associations over a range of physical scales with a watershed algorithm.  We describe the cross-observatory imaging, catalogs, and software products to be released. These high-level science products will seed a broad range of investigations, in particular, the study of embedded stellar populations and dust with JWST, for which a PHANGS Cycle 1 Treasury program to obtain eight band 2--21 $\mu$m imaging has been approved. 

\end{abstract}

\keywords{star formation --- star clusters --- spiral galaxies --- surveys}

\section{Introduction}
\label{sec:intro}

How do stars form from the complex multi-phase interstellar medium (ISM) in galaxies?  This question lies at the heart of astrophysics, as star formation is a key mechanism governing the evolution of baryons in the universe \cite[e.g.,][]{peroux2020}.  Star formation converts interstellar matter into stars and their planetary systems, depletes galaxies of gas, and feeds back metals, energy, and momentum into the ISM, which may reach the halos of galaxies and beyond.  In turn, this feedback, together with galactic-scale inflows and dynamics, impacts the state of the gas and the future of star formation.

Many processes underlie this star formation cycle, and nearly all have been the focus of dedicated study.  Star formation typically occurs in molecular clouds \cite[e.g.,][]{blitz1993,heyer2015,miville-deschenes2017}, parts of which can become gravitationally unstable and contract until new stars are born \cite[e.g.][]{krumholz2005,dobbs2014,chevance2020}. This process is controlled by the intricate interplay between self-gravity and various opposing agents, such as supersonic turbulence, magnetic fields, radiation, and gas and cosmic ray pressure \cite[e.g.,][]{bruce00, mckee07, girichidis2020}. 
The flow patterns within the galactic ISM help determine where and at what rate stars form, and are themselves influenced by the energy and momentum input from massive stars \cite[e.g.,][]{maclow2004,mckee07,hennebelle2012,federrath2013}. That is, the local process of stellar birth is 
impacted by the supply, organization, and stability of cloud-scale natal gas as governed by large-scale galaxy dynamics, including spiral arm features or perturbations from satellite galaxies or accretion of fresh gas from the cosmic web \cite[e.g.][]{kennicutt98,dobbs06,dobbs08, leroy2008,leroy2013,meidt13, meidt20}. 
Stellar feedback, in the form of radiation, winds, and supernova explosions, creates a hierarchy of highly non-linear feedback loops which impacts ISM dynamics across a wide range of physical scales \cite[e.g.][]{hopkins2014,lopez2014,walch2015,gnedin2016,rahner2017,olivier21}, thereby determining the chemical and thermal state of the ISM, and affecting subsequent star formation \cite[e.g.][]{klessen2016}. 

We now understand from decades of study and observations across the electromagnetic spectrum that all of these processes that drive, regulate and extinguish star formation operate together over a vast range of stellar, interstellar, galactic, and circumgalactic scales.  Accordingly, we have come to recognize that systematic observations -- spanning essential spatial scales and phases of the star formation cycle, across different galactic environments -- are necessary for the development of a robust, unified model of star formation and galaxy evolution.  

Here, we present the PHANGS-HST Treasury survey, which as part of the Physics at High Angular Resolution in Nearby Galaxies\footnote{\url{http://www.phangs.org}} (PHANGS) program, is building a dataset for the systematic multi-scale, multi-phase study of star formation.  PHANGS is charting the connections between giant molecular clouds, {\sc Hii} regions, and young stars throughout a diversity of galactic environments in the local universe by combining observations from large surveys with ALMA, VLT/MUSE, and HST.  Supporting data including VLA {\sc Hi}, and Astrosat FUV/NUV imaging, as well as the wealth of panchromatic ground- and space-based survey observations obtained for the nearby galaxies in the sample over the past three decades, have also been assembled.  There are currently three major components of PHANGS:

\textbf{PHANGS-ALMA}: The foundation of PHANGS has been built with the transformative capabilities of the Atacama Large Millimeter/\linebreak[0]{}submillimeter Array (ALMA). Through a Cycle~5 PHANGS-ALMA Large Program (PI E.~Schinnerer) and smaller precursor programs, PHANGS has obtained $\sim$1\arcsec\ resolution \mbox{CO(2--1)} maps for a sample of 74 massive spiral galaxies at distances of 4--23 Mpc.  At these distances, ALMA can detect individual giant molecular clouds with better than 2.4 km~s$^{-1}$ velocity resolution and physical resolutions of ${\sim}50{-}100$~pc, while still efficiently covering the star-forming disk \citep[][]{sun18,phangs-alma}.

\textbf{PHANGS-MUSE \& PHANGS-H$\alpha$}: For 19 of these galaxies, with the Very Large Telescope/\linebreak[0]{}Multi Unit Spectroscopic Explorer, PHANGS-MUSE (PI E.~Schinnerer) has obtained IFU spectroscopy with $\sim$2.5~\AA\ spectral resolution and $\sim$0\farcs7 spatial resolution to deliver a 3D view of the ionized (10$^4$~K) gas, stellar populations, and kinematics via various gas and stellar tracers in the optical from 4800--9300~\AA\ \citep[E.~Emsellem et al. submitted; see first results in][]{kreckel18, ho19, kreckel19, kreckel20}.  To supplement the MUSE observations, the PHANGS-H$\alpha$ survey (A.~Razza et al. in preparation) has obtained seeing-limited (${\sim}1\arcsec$) narrow-band H$\alpha$ imaging, to provide star formation rate (SFR) maps and catalogs of ionized nebulae for the full PHANGS-ALMA sample. The H$\alpha$ imaging was obtained using WFI on the ESO/MPG 2.2m telescope at La~Silla, and the DirectCCD on the du~Pont 2.5m telescope at Las Campanas Observatory. 

\textbf{PHANGS-HST}, the subject of this paper, is a Cycle~26 Hubble Space Telescope Treasury survey (PI J.~C.~Lee) which has obtained $NUV$-$U$-$B$-$V$-$I$ imaging of the disks of \ngal\ galaxies from the parent PHANGS-ALMA sample, and parallel V and I band imaging of their halos.  The HST sample includes all 19 galaxies with MUSE IFU spectroscopy.  


The high-resolution capabilities of HST ($\sim$0\farcs08) have enabled the study of compact star clusters and associations in galaxies out to distances of several tens of Mpc \citep[e.g.,][]{whitmore99, linden17,adamo20}.  These structures, which typically have half-light radii of a few parsec \citep{pz10,ryon17,krumholz19}, have been the focus of much recent work and are not only important to study in their own right \citep[e.g.,][]{whitmore07, chandar10a, kruijssen12, krumholz19, adamo20},
but also have great utility as `clocks' -- effectively single-aged stellar populations that can be age-dated and used to time various star formation and ISM processes.  The PHANGS-HST UV-optical imaging enable inventories of young star clusters and associations down to a few thousand solar masses, with age and mass determinations from SED-fitting accurate to a factor of $\sim$2 \citep[e.g.][]{turner21} on average. 

Altogether, PHANGS will yield an unprecedented sample of $\sim$100,000 star clusters, associations, {\sc Hii} regions, and molecular clouds in diverse galactic environments to provide answers to the following open questions:
\begin{itemize}
    \item What are the timescales for the onset of star formation in clouds, the destruction of clouds, and the removal of gas from young star clusters?
    \item How are the mass functions of star clusters/\linebreak[0]{}associations related to those of clouds? What are the implied star formation efficiencies?
    \item How are star formation and gas organized into multi-scale structures? How do their relative spatial distributions evolve with time?
\end{itemize} 

These questions, particularly those examining the relationship between molecular clouds and star clusters, have been posed in the context of the Milky Way (Murray 2011; Lee et al. 2016) and select Local Group and Nearby galaxies (e.g., M51: Hughes et al. 2013, Grasha et al. 2019; NGC7793: Grasha et al. 2018; NGC300: Kruijssen et al. 2019). But whether the answers to the questions vary with galactic environment is still unclear, as there has not yet been a systematic study on the cluster scale across a well-defined sample of galaxies spanning a broad range of global properties.
PHANGS will provide not only the answers to these questions, but moreover how they depend on galactic properties such as the phase balance and physical conditions of the ISM, stellar mass, gas mass, SFR (as well as their surface densities), metallicity, and the presence of dynamical features such as rings, bars, and spiral arms.

Here, we present the design of the PHANGS-HST survey, and provide an overview of the data processing pipeline developed to generate the data products required for the investigation of these questions.  The remainder of this paper is organized as follows.  In Section~\ref{sec:sample}, we describe how galaxies are selected from the parent PHANGS sample for observation with HST, and summarize the global properties of the sample.  Our HST imaging observations with WFC3 and ACS are described in Section~\ref{sec:observations}.  Section~\ref{sec:pipeline} describes the PHANGS-HST pipeline used to produce catalogs of compact star clusters and associations, and to measure their observed and physical properties.  This high-level description is intended to provide the framework for a series of papers which document each of the major components in detail, in particular new methods for selecting star cluster candidates, morphological classification of candidates, and identification of stellar associations over a range of physical scales.  Data products resulting from this pipeline which will be released to support community science are described in Section~\ref{sec:dataproducts}.   In Section~\ref{sec:summary}, we conclude with a summary, and look ahead to upcoming work on dust and young embedded stellar populations with JWST.  Through a Cycle 1 Treasury program, we will add a fourth major component of PHANGS and obtain eight band 2--21 $\mu$m imaging for the 19 galaxies with the full suite of PHANGS ALMA, MUSE, and HST observations.

Magnitudes in this and other PHANGS-HST pipeline papers are given in the Vega system, unless otherwise noted, to facilitate comparison with prior HST studies of resolved stellar populations.



\section{Galaxy Sample}
\label{sec:sample}
\begin{figure*}
    \centering
    \includegraphics[width=2.85in]{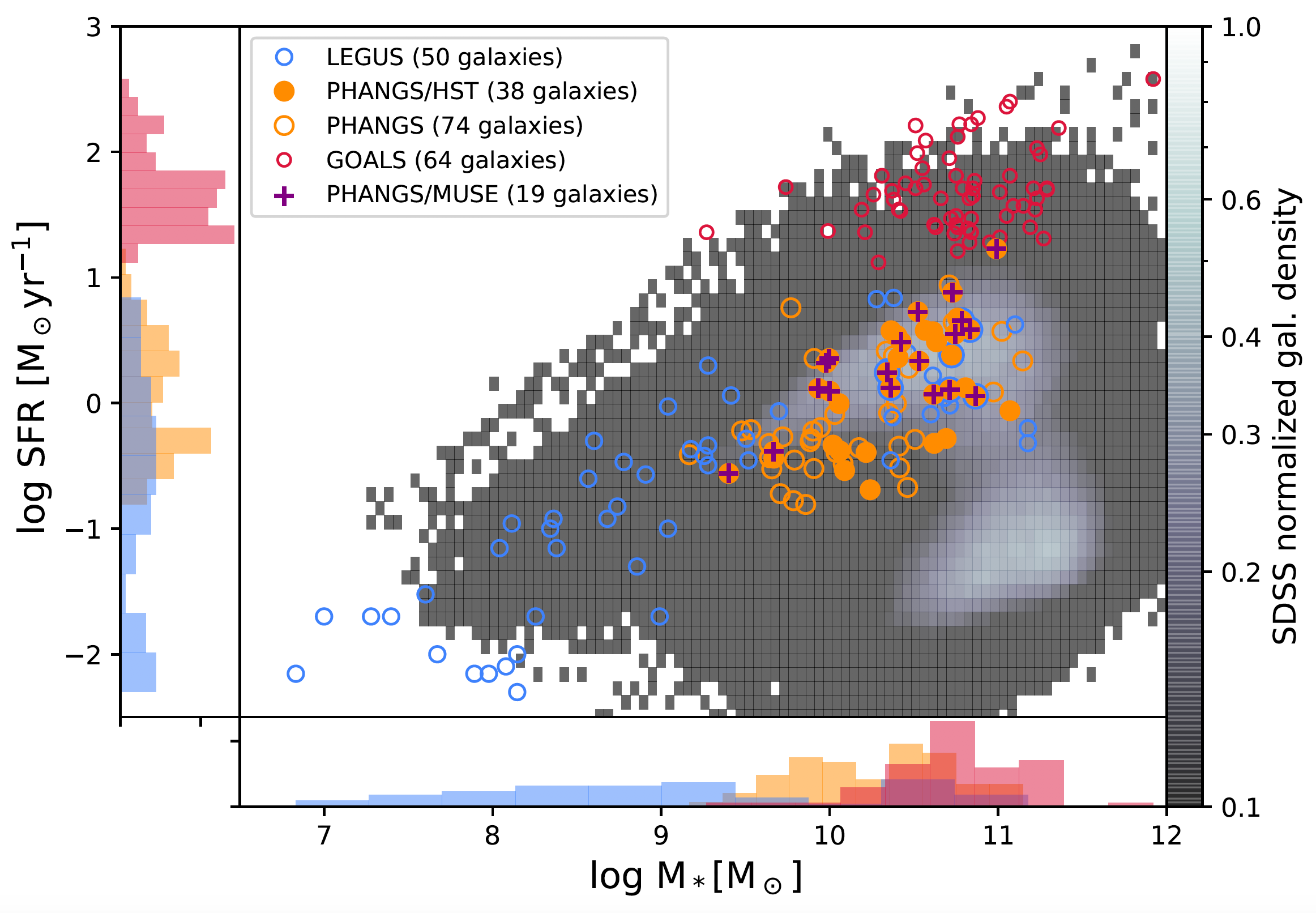}
    \includegraphics[width=2.85in]{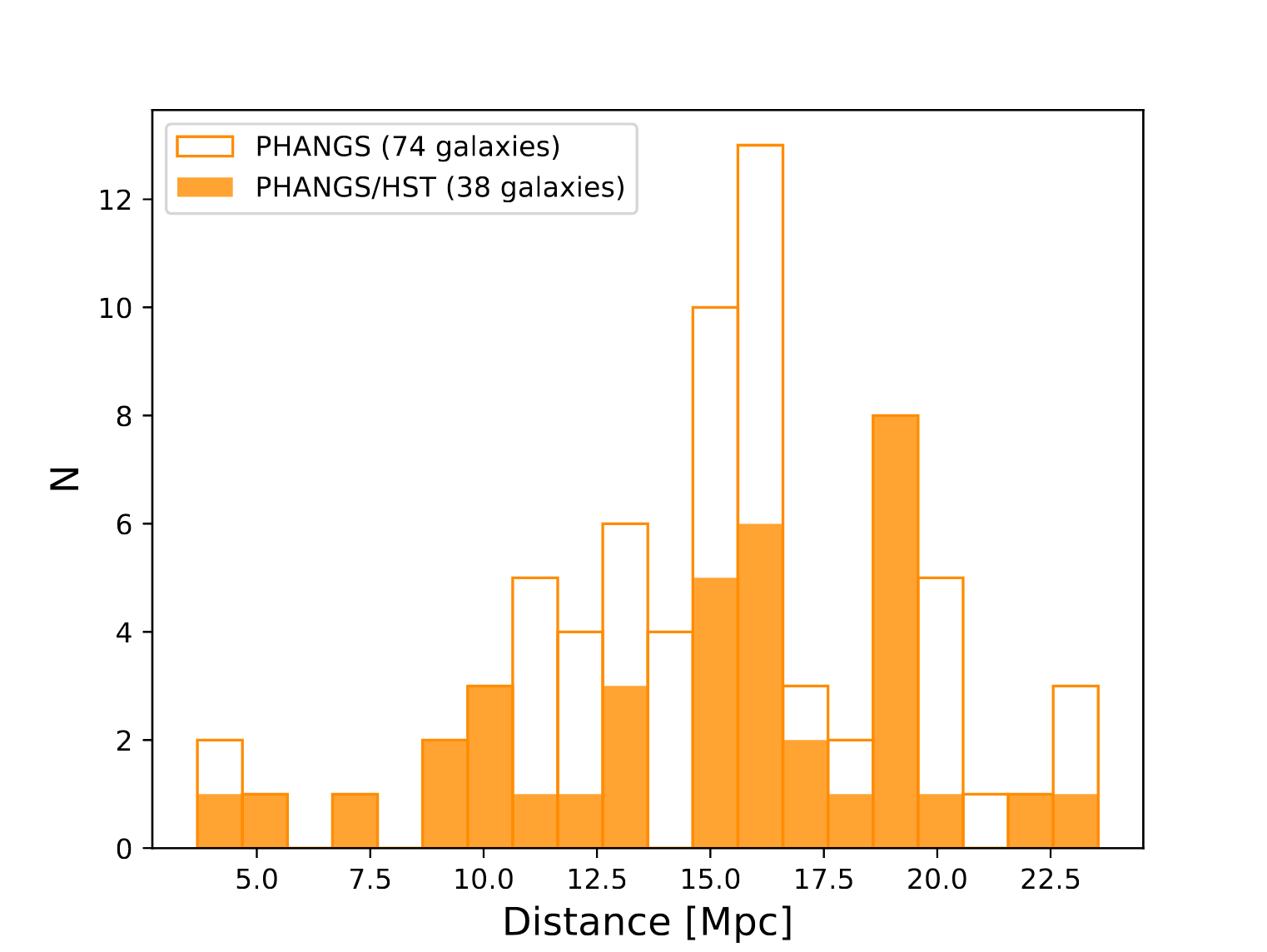}
    \includegraphics[width=2.85in]{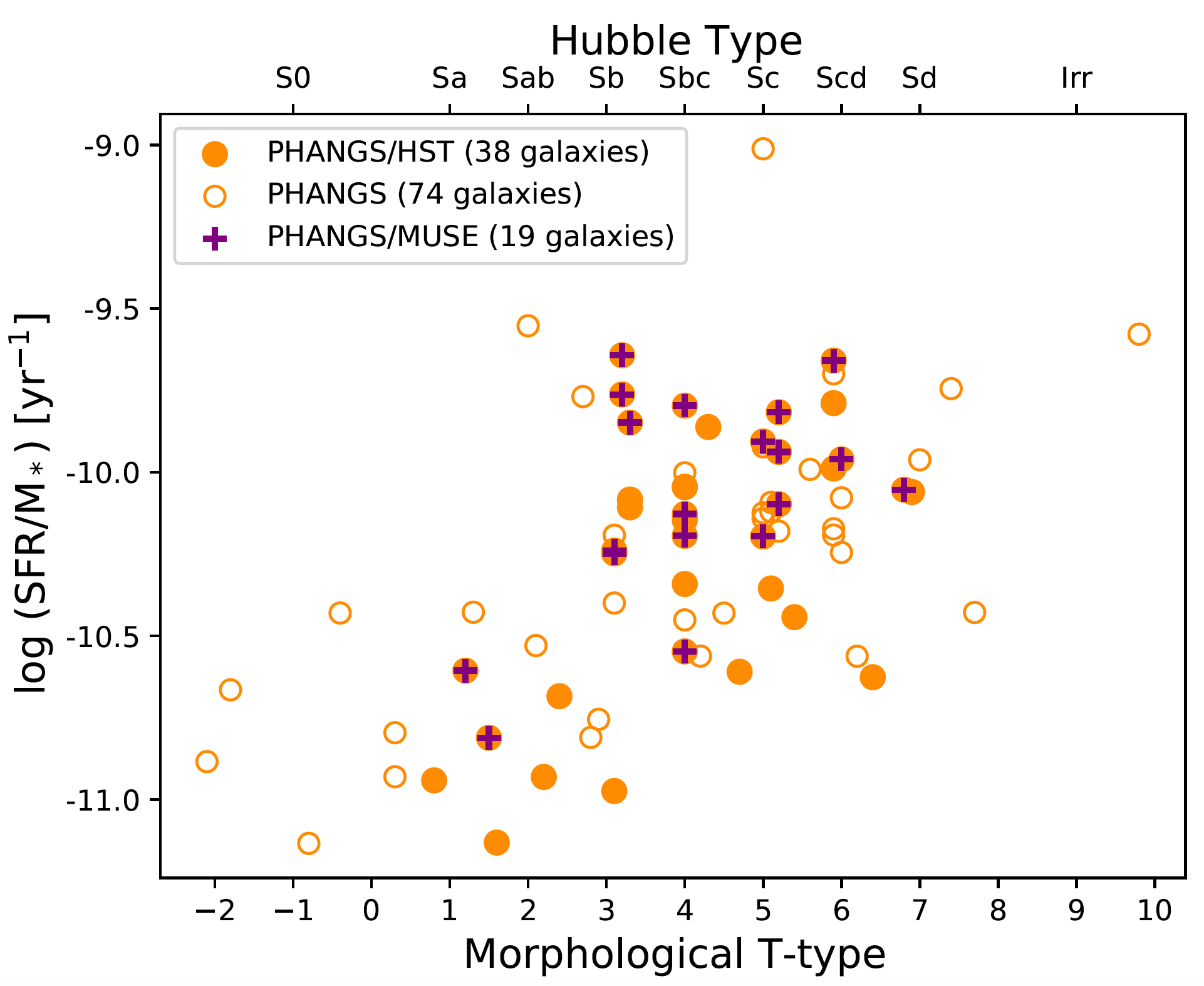}
    \includegraphics[width=2.85in]{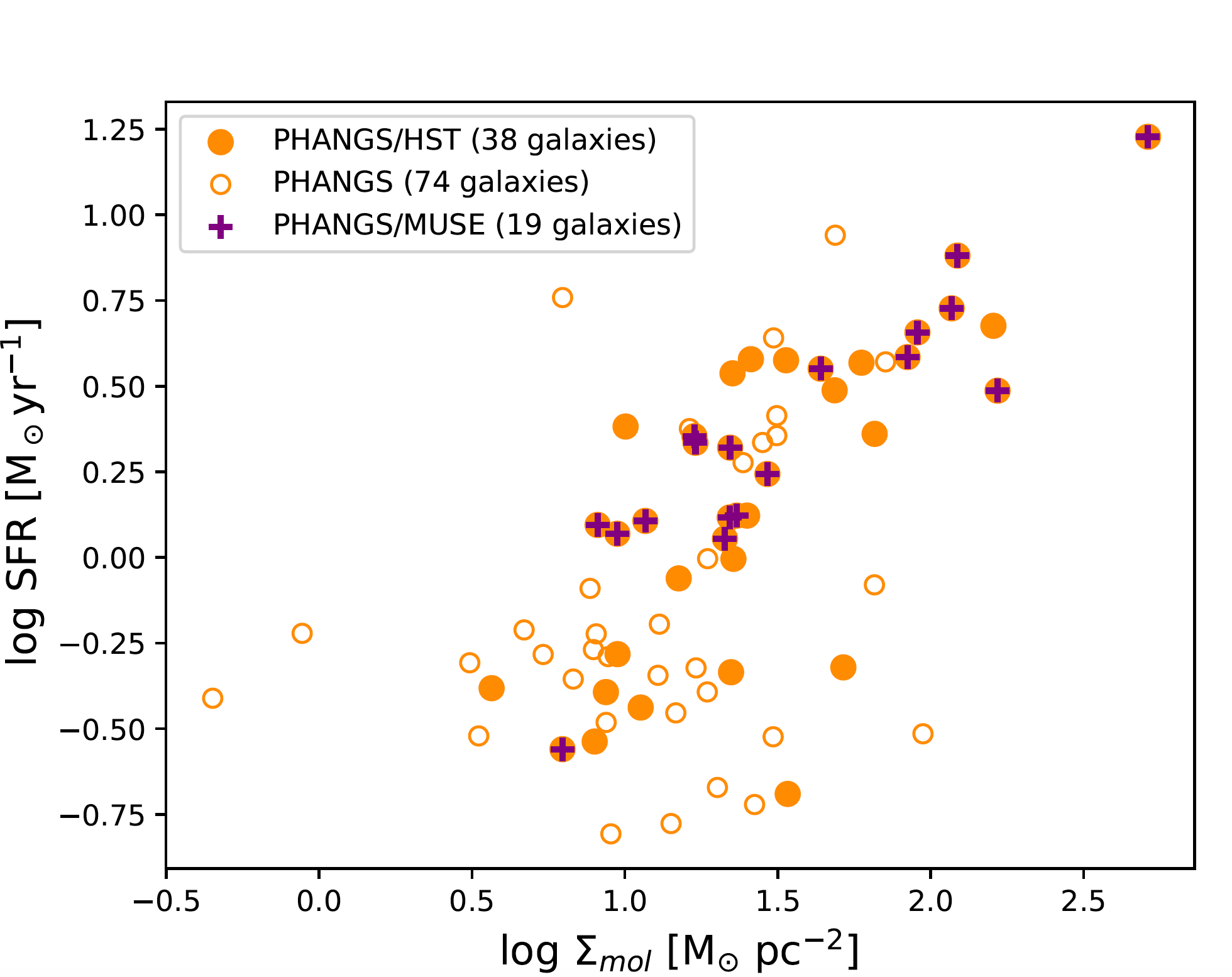}
    \caption{\textbf{Upper left:} Coverage of the SFR--M$_*$ plane by galaxies in the parent PHANGS sample ($N=74$; orange open and filled points), and those targeted for HST observations ($N=\ngal$; orange filled points).  The PHANGS-HST sub-sample is representative of the $d \lesssim 20$~Mpc massive galaxy population on the local star-forming main sequence, which contain the bulk of molecular gas and present-day star formation.  Shown for context are: an SDSS local galaxy sample, which extends over a much larger volume to $z \lesssim 0.3$ \citep[greyscale;][]{salim16}; the HST LEGUS sample \citep[blue;][]{legus}, which focuses on the nearest galaxies ($d \lesssim 11$~Mpc) and includes a significant number of lower mass dwarfs; and luminous infrared galaxies within $\sim$350 Mpc from the GOALS sample \citep[red;][]{u12}. \textbf{Upper right:} Distribution of distances of galaxies in the PHANGS parent sample, and those selected for HST observations.  All distances are provided in Table \ref{tab:galaxysample}, and are taken from the compilation of \cite{anand20}, with an update for one new TRGB distance derived from PHANGS-HST observations taken after the publication of that paper (see Sec~\ref{sec:parallel}). Altogether, the PHANGS-HST V and I band parallel imaging  have yielded 8 TRGB measurements, which represent the best available distances for those galaxies. \textbf{Bottom panels:} PHANGS-HST also provides coverage of a full range of specific SFR, molecular gas surface densities, and spiral galaxy morphologies.  Galaxies with MUSE IFU spectroscopy are indicated ($N=19$) with crosses in the upper left and bottom plots.  These 19 galaxies comprise the sample targeted by the PHANGS-JWST Cycle 1 Treasury program (see Sec~\ref{sec:summary}).} 
    \label{fig:sample}
\end{figure*}

Galaxies are chosen for HST observations from the PHANGS-ALMA Large Program sample (PI E. Schinnerer). PHANGS-ALMA has obtained \mbox{CO(2--1)} maps for a complete sample of \ngalphangs\ southern galaxies\footnote{The original PHANGS-ALMA Large Program sample, from which PHANGS-HST targets were selected, consists  of \ngalphangs\ galaxies.  PHANGS has now extended the sample to 90 galaxies to include additional nearby galaxies with CO mapping available from the ALMA archive, as well as early-type galaxies, as explained in more detail in \citet{phangs-alma}.} ($-75^{\circ}<\delta<20^{\circ}$; i.e. ALMA-observable), which were selected to be massive ($M_{*} \gtrsim 10^{9.75}$~\msun), star-forming, not edge-on to the line-of-sight, and at distances $\lesssim$17~Mpc.\footnote{After a recent update to the distance determinations, including the addition of new TRGB distances from our parallel ACS V and I band imaging (Section~\ref{sec:parallel}), we find that galaxies in the PHANGS-HST sample lie at distances between 4.4 and $\sim$23 Mpc, with a median of $\sim$16 Mpc (Figure~\ref{fig:sample}).  Uncertainties in distances that were used in the initial PHANGS galaxy selection, some of which were based on recessional velocities corrected for peculiar motions based on a flow model, led to the inclusion of galaxies which lie beyond the initial $\sim$17~Mpc limit.  Further discussion of the impact of distance uncertainties on the PHANGS sample selection can be found in \citet{phangs-alma}. Full details on the compilation of best available distances are provided in \cite{anand20}, and summarized in Section~\ref{sec:parallel}.}  
\mbox{CO(2--1)} observations were obtained for this sample via an ALMA Cycle~5 Large Program (2017.1.00886.L), which builds upon and incorporates several smaller precursor programs in Cycles 2--3.\footnote{Cycle~2 program 2013.1.00650.S (PI E.~Schinnerer), Cycle~3 program 2013.1.01161.S (PI K.~Sakamoto), Cycle~3 program 2015.1.00925.S (PI G.~A.~Blanc), Cycle~3 program 2015.1.00956.S (PI A.~K.~Leroy) and additional programs in Table~2, \citet{phangs-alma}.}  A full description of the PHANGS-ALMA Large Program sample criteria, and derivation of integrated properties used for selection, such as stellar mass ($M_*$), SFR, and integrated CO luminosities, is given in \citet{phangs-alma}.

With HST, we target the galaxies best-suited for joint HST-ALMA analysis of resolved young stellar populations and giant molecular clouds. That is, we select galaxies from the PHANGS-ALMA parent sample that (1) have inclinations $i \lesssim 70^{\circ}$, to minimize source blending and attenuation along the line-of-sight due to dust within the target, (2) avoid the Galactic plane ($|b|>15^{\circ}$), to minimize the impact of Milky Way reddening and foreground stars, and (3) are sufficiently active (SFR $\gtrsim$ 0.3 \sfr), to ensure that wide-spread molecular cloud and star cluster populations are available for joint study.  The resulting set of \ngal\ galaxies chosen for HST observations is given in Table~\ref{tab:galaxysample} along with basic properties (which have been refined and updated since the sample was first compiled) relevant to their selection.  The 19 galaxies for which VLT/MUSE optical integral field spectroscopy has been obtained are all included by the HST selection criteria.

The \ngal\ PHANGS-HST galaxies probes a full range of global properties covered by the PHANGS-ALMA parent sample, which is itself representative of the overall present-day spiral galaxy population.  This is illustrated in Figure~\ref{fig:sample} (upper left panel), which shows $M_*$ and SFR of the PHANGS sample and the subset targeted with HST, overlaid on the locus occupied by star-forming galaxies from the Sloan Digital Sky Survey.  
Figure~\ref{fig:sample} shows that the PHANGS sample provide excellent coverage of the galaxy “main sequence” between stellar masses of ${\sim}10^{9.5} {-} 10^{11}$~\msun.  Main sequence galaxies in this stellar mass range are representative of the environments where the bulk of molecular gas and present-day star formation are found \citep{salim07, Saintonge17}.  The bottom panels of Figures~\ref{fig:sample} illustrate the coverage of specific SFR as a function of morphological type and the SFR as a function of the molecular gas surface density.  The PHANGS-HST observations include spiral galaxies with morphological types of Sa through Sd, sSFRs from ${\sim}10^{-10.5} {-} 10^{-9}$~yr$^{-1}$, SFR from ${\sim}0.2{-}17$ \sfr, and $\Sigma_{\mbox{mol}}$ from ${\sim}10^{0.5}{-}10^{2.7}$ \msun~pc$^{-2}$.

To place the PHANGS-HST survey in further context, nearby galaxies from two other complementary HST imaging programs are also shown in the upper left panel of Figure~\ref{fig:sample}: HST LEGUS \cite[Legacy ExtraGalactic Ultraviolet Survey;][]{calzetti15}, and GOALS \cite[Great Observatories All-Sky LIRG Survey;][]{armus09}.  As a Cycle~21 Treasury program, LEGUS also obtained 5-band UV-optical imaging to study star clusters in a representative sample of 50 galaxies \citep{adamo17}, but focused on the nearest late-type systems \citep[$\lesssim$11 Mpc;][]{rck08,jcl11} to also enable the re-construction of star formation histories from individually resolved stars \citep{cignoni18,sacchi18,cignoni19}.  LEGUS is therefore naturally dominated by lower mass, local volume dwarf and irregular galaxies \citep[e.g.,][]{lee09a}, which comprise about half of its 50 galaxy sample \citep{cook19}.  PHANGS-HST builds on the groundwork laid by programs such as LEGUS, and earlier transformative work by the Panchromatic Hubble Andromeda Treasury \citep[PHAT][]{phat} and the WFC3 Early Release Science (ERS) program\footnote{\url{https://archive.stsci.edu/prepds/wfc3ers/}. The WFC3 imaging observations obtained by these successive programs, which provide multi-wavelength coverage beginning in the near-UV (with ERS using up to 11 filters including both broad and narrow bands), have enabled the study of the physical properties of star clusters in samples of main sequence galaxies that have grown from a few to over one hundred in the past decade \citep[e.g.,][]{chandar10b,chandar14,johnson16,johnson17,adamo17,grasha17,cook19}.}  At the edge of the parameter space occupied by nearby star-forming galaxies, GOALS has obtained HST imaging in the B, I, and H filters for $\sim$90 luminous infrared galaxies with thermal IR ($8{-}1000~\mu$m) dust emission greater than  $10^{11}$~L$_{\odot}$ \citep{haan11,kim13}.  (A subsample of 64 LIRGS from \cite{u12} with stellar masses derived from SED fitting is shown in Figure~\ref{fig:sample}.) Such highly active star-forming galaxies are rare in the present-day universe, so the GOALS sample extends over much larger distances compared to LEGUS and PHANGS-HST.  GOALS galaxies are located at distances of up to $\sim$350~Mpc ($z\lesssim0.08$), and hence studies of the stellar populations have focused on larger ``clumps'' ($\sim$90 pc, \citealt{larson20}, but also see \citealt{linden17}).  Currently PHANGS is the only program with uniform ALMA CO observations for a significant sample of nearby galaxies, but ultimately, analysis of all of these programs together are needed to fully understand the impact of galactic environment on star formation.




\section{HST Observations}
\label{sec:observations}
\begin{figure}[t!]
\includegraphics[width=\linewidth]{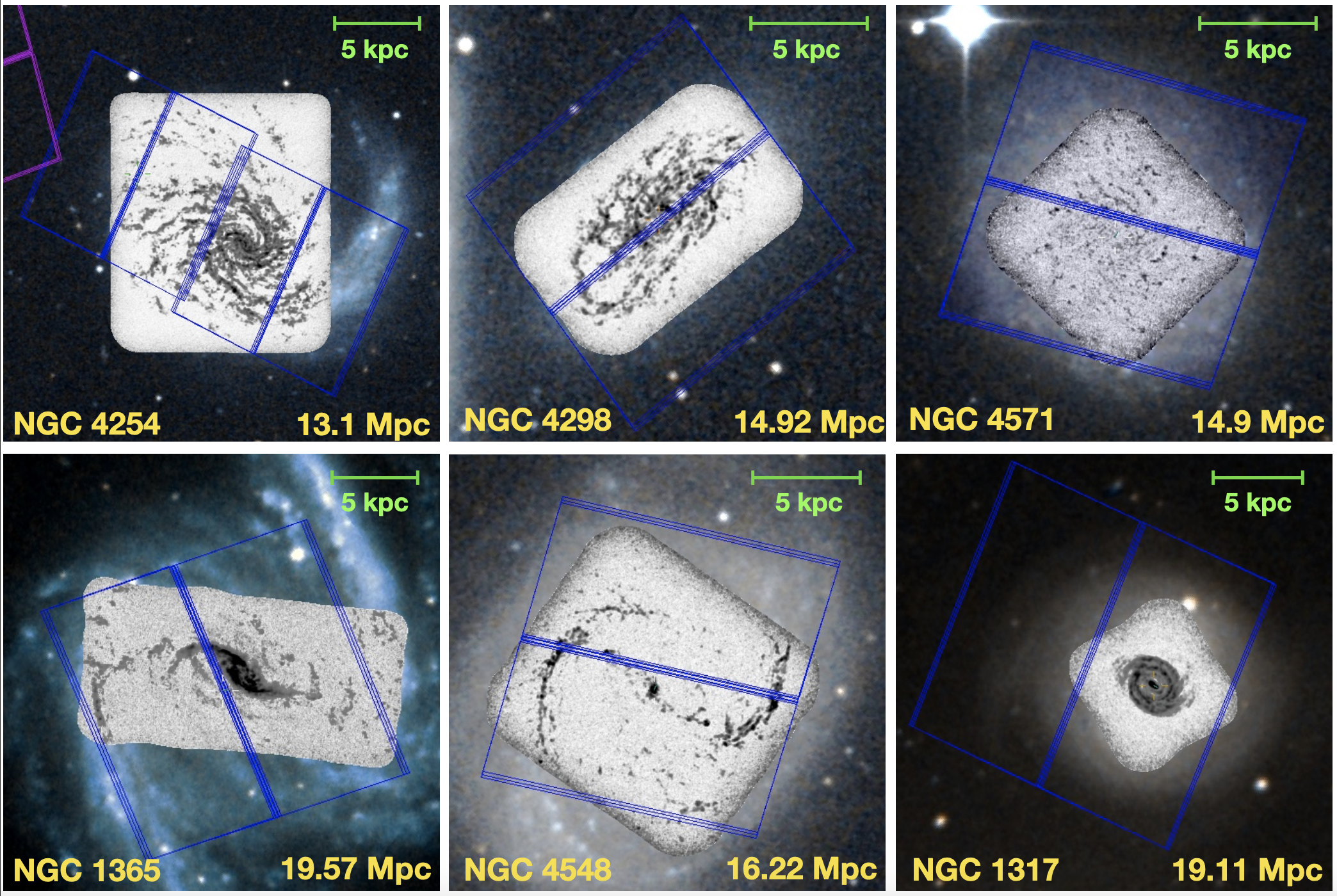}
 \caption{PHANGS-HST WFC3 UVIS footprints (162\arcsec$\times$162\arcsec) overlaid on PHANGS \mbox{CO(2--1)} ALMA maps for 6 galaxies in our sample of 38, showing the enormous diversity  of  molecular  gas  content  and  morphology  in  present-day  massive  star-forming  galaxies.    CO maps are overlaid on wider field DSS imaging.  Scale bars and galaxy distance are shown in the upper and lower right corners respectively. \textbf{Top:}  Targets showing decreasing molecular gas surface  density and specific SFR, from left to  right. \textbf{Bottom:} Impact of dynamical features on the gas distribution; two examples showing differing responses of the  gas  to  the  influence of a bar  (left,  middle)  and an  example  of  a  ring  feature  (right).  Our  WFC3/UVIS observations are allowing us to find and characterize young stellar clusters and associations over the same area  covered  by these detailed CO maps, to create the  first  combined  atlas  of clouds  and clusters across a representative sample of massive main sequence galaxies in the local universe. }
 \label{fig:footprints}
\end{figure}

\begin{figure}[t!]
\includegraphics[width=\linewidth]{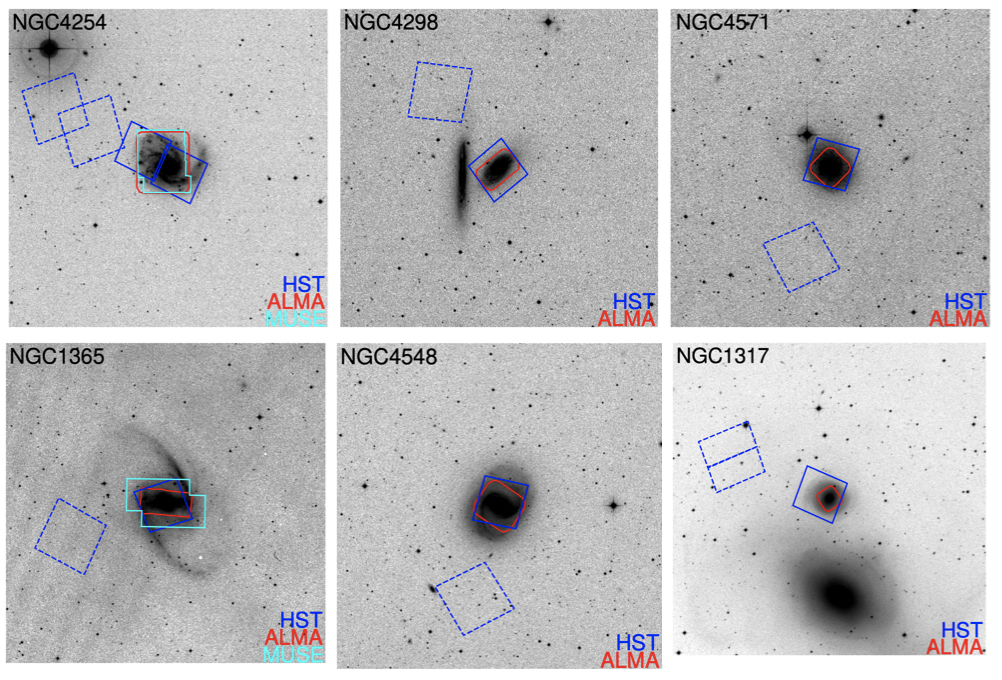}
 \caption{Figures showing the overlap of the PHANGS-HST WFC3 UVIS (blue), ALMA (red), and MUSE (where available; cyan) observation footprints, overlaid on DSS imaging for the same 6 galaxies as in Figure \ref{fig:footprints}.  A larger field (20\arcmin$\times$20\arcmin) is shown relative to Figure \ref{fig:footprints} to illustrate the placement of the HST ACS parallel pointing (dashed lines).  The WFC3 UVIS field-of-view is 162\arcsec$\times$162\arcsec and the ACS field-of-view is 202\arcsec$\times$202\arcsec.  Such footprint maps for the full PHANGS-HST sample can be found at \url{https://archive.stsci.edu/hlsp/phangs-hst/}.}
 \label{fig:footprints2}
\end{figure}

\begin{figure*}
\includegraphics[width=\textwidth]{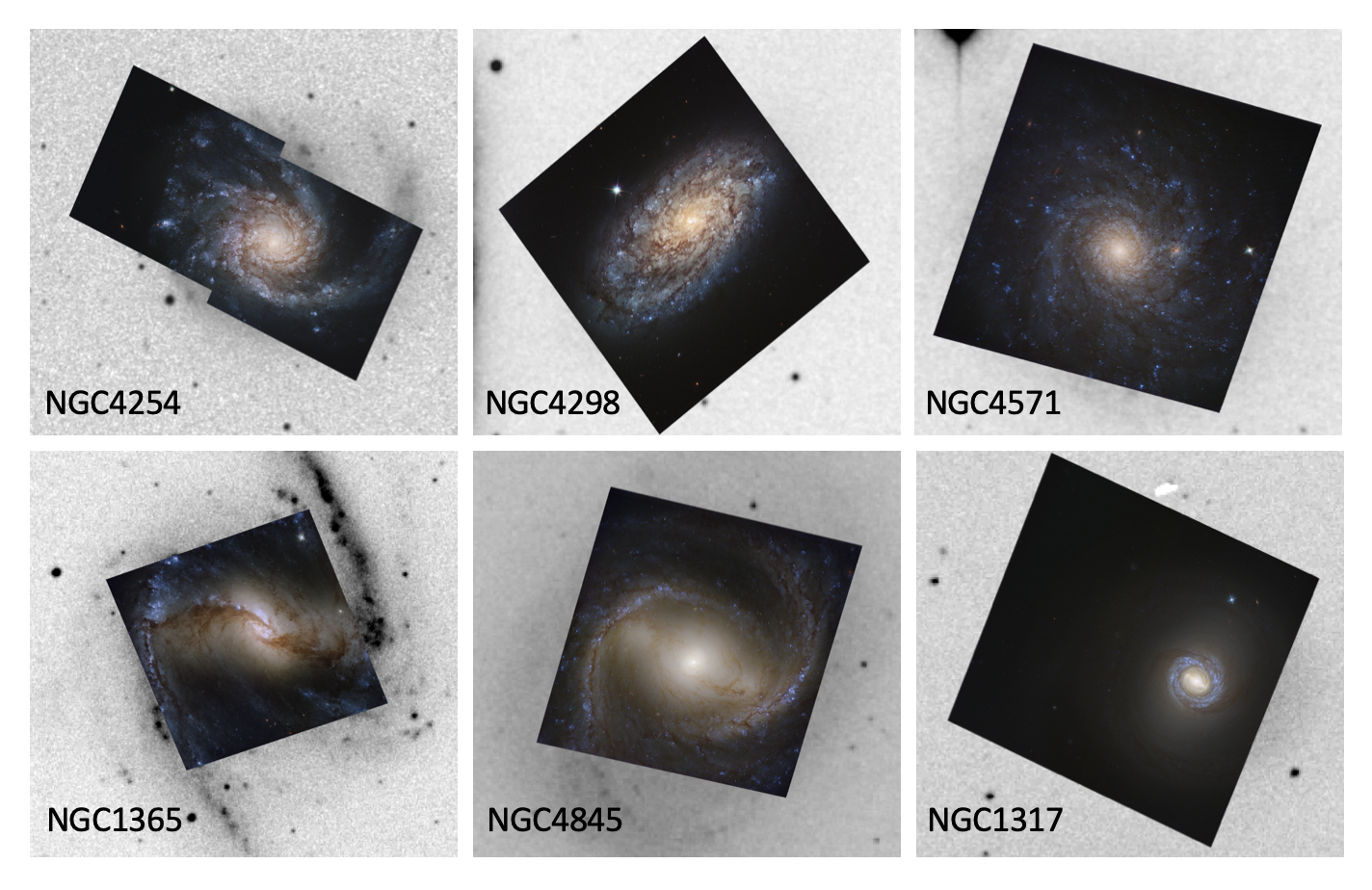}
 \caption{Color composites of PHANGS-HST imaging (Red: WFC3/UVIS F814W,
Green: WFC3/UVIS F555W, Blue: WFC3/UVIS F438W+F336W+F275W), overlaid on DSS imaging for the same 6 galaxies as in Figures \ref{fig:footprints} and \ref{fig:footprints2}. PHANGS-HST imaging for nine galaxies have been featured as the ESA/Hubble image of the week (\url{https://phangs.stsci.edu/\#news}).}
 \label{fig:footprints3}
\end{figure*}

Imaging observations for the PHANGS-HST Treasury program (Cycle~26, PID 15654) were conducted from April 2019 to April 2021 with an allocation of 122 orbits.  Previous to this program, no HST wide-field UV imaging was available for 80\% of the PHANGS-HST sample, and 60\% also did not have any optical imaging with either WFC3 or ACS.  Thus, PHANGS-HST provides a critical augmentation to the HST archive for the set of nearby spiral galaxies where both star clusters and molecular clouds can be efficiently detected by HST and ALMA over galactic scales.

As discussed in Section~\ref{sec:sample}, the PHANGS-HST sample contains a total of \ngal\ galaxies.  
No new observations were conducted for 4 of these galaxies, since sufficient imaging for those targets were previously obtained by the LEGUS program.\footnote{NGC~0628, NGC~1433, NGC~1512, and NGC~1566. Another three galaxies in the PHANGS-HST sample also have observations available from LEGUS (NGC~3351, NGC~3627, NGC~6744), but additional observations were obtained to increase the coverage of the area of the disk mapped by ALMA in \mbox{CO(2--1)}.}  Altogether, new observations were planned for 43 fields in 34 galaxies.  
Initial observations for 8 fields were corrupted due to guiding failures, but were all rescheduled and successfully observed on the second attempt.  These re-observations were the primary reason the execution of the program spanned over two years instead of one.

To illustrate the PHANGS coverage of each galaxy, footprints of the HST and ALMA observations, together with those for VLT/MUSE when available, are provided at MAST at \url{ https://archive.stsci.edu/hlsp/phangs-hst/}.  Examples of these footprint overlays are given in Figures~\ref{fig:footprints} and~\ref{fig:footprints2} for six galaxies chosen to span the range of molecular gas surface densities, specific SFRs, and dynamical features (rings and bars) present in the sample.  Figure~\ref{fig:footprints} focuses on the PHANGS-HST prime imaging area in the context of the PHANGS-ALMA \mbox{CO(2--1)} maps.  Figure~\ref{fig:footprints2} shows a wider areal view of each galaxy with Digitized Sky Survey images,\footnote{The Digitized Sky Surveys were produced at the Space Telescope Science Institute under U.S. Government grant NAG W-2166. The images of these surveys are based on photographic data obtained using the Oschin Schmidt Telescope on Palomar Mountain and the UK Schmidt Telescope. The plates were processed into the present compressed digital form with the permission of these institutions.} and also includes footprints of the PHANGS-HST parallel, PHANGS-ALMA, and PHANGS-MUSE observations.


\subsection{Prime Observations with WFC3}
For each of the 43 new fields, we aimed to cover the region mapped in \mbox{CO(2--1)} by ALMA in five filters: F275W (NUV), F336W (U), F438W (B), F555W (V), F814W (I).  F275W is the shortest wavelength filter that avoids the 2175~\AA\ dust feature, and in combination with the U and B bands serves to break the age-extinction degeneracy.  The V and I bands are less affected by extinction and variation in the mass-to-light ratio, and serve to constrain the stellar mass.  

For 31 fields (in 26 galaxies), new observations with the WFC3 UVIS camera were needed in all five filters.  Three exposures with sub-pixel dithering were taken in each filter (the average pixel size of WFC3 UVIS is 0\farcs04), and all 15 exposures were obtained in a 3-orbit visit, yielding total exposure times of $\sim$2200~s (NUV), $\sim$1100~s (U), $\sim$1100~s (B), $\sim$670~s (V), $\sim$830~s (I).  A post flash of 5--10~e$^{-}$ is applied for each exposure, with larger values in the shorter wavelength filters, to increase the background to 12~e$^{-}$, which is the recommended value to mitigate issues due to charge transfer efficiency (CTE) losses.  The dither sequence is optimized to cover the WFC3 chip gap and to help recover the under-sampled point spread function (PSF). 

For the other 12 pointings (in 8 galaxies), suitable WFC3 or ACS data in one or more of these filters were taken by prior programs.  The available archival data were obtained from MAST and processed in a consistent manner with our new observations, and new imaging was obtained only for the missing filters within a 2-orbit visit.


\subsection{Parallel Observations with ACS}
\label{sec:parallel}
While the primary observational goal of the PHANGS-HST survey is to obtain UV-optical imaging of the star-forming disk, we also simultaneously observe the galaxy halo with the Advanced Camera for Surveys Wide Field Channel (ACS/WFC) in ``parallel'' mode.  Such parallel observations can potentially yield measurements of the galaxy distance if the tip of the red giant branch (TRGB) can be identified in a color-magnitude diagram of halo stars.  Thus, we have designed our observations so that ACS imaging in F606W and F814W, filters commonly used for TRGB analysis \citep[e.g.,][]{mcquinn17,anand18} accompanies each WFC3 ``prime'' observation.  

For the range of distances and angular sizes of the PHANGS-HST galaxies, the ACS field-of-view generally falls on the halo of the target galaxy when WFC3 is centered on the galaxy itself (Figure~\ref{fig:footprints2}). Given that the science requirements of PHANGS-HST constrain the positioning of the prime pointings, optimizing placement of the parallel fields (as in a focused TRGB program) is a secondary priority and is restricted by the fixed spatial offset of the two cameras on the focal plane.  For galaxies with relatively large angular sizes, the parallel observations may include portions of the outer disk.  For smaller galaxies, the parallels may be too far to capture a significant number of halo stars.  To the extent possible, ORIENT constraints were imposed to prevent the ACS field from entirely falling on the galaxy disk, on nearby galaxy neighbors, and/or on extremely bright foreground stars.  For some targets with large angular sizes where it was not possible to avoid the disk completely, the field was positioned along the major axis to help differentiate between disk and halo stars. In several cases, the desired ORIENT constraints were lifted or relaxed to allow guide stars to fall into the area of the focal plane accessible to the Fine Guidance Sensors. 

The five-band prime observations with WFC3/UVIS were sequenced in each orbit to optimize exposure time in the parallel observations without impacting the primary observations.  As discussed in the previous section, WFC3 observations for each pointing required 2 or 3 orbits, depending on whether suitable HST archival observations of the galaxy disk were already available.  For 2-orbit visit pointings, the total exposure times in the ACS parallel V and I images are  $\sim$2100~s each, while for the 3-orbit visits they are about $\sim$3500~s and 3200~s, respectively.  In both cases, three exposures were taken in each filter.  

Distance constraints resulting from analysis of the TRGB based on parallel imaging obtained in the first year of the PHANGS-HST program (for 30 galaxies through 2020 July) are presented in \cite{anand20}.  We were able to measure TRGB distances for 10 PHANGS galaxies, 4 of which are the first published TRGB distance measurements for those galaxies (IC 5332, NGC 2835, NGC 4298, NGC 4321) and 7 of which represent the best-available distances (IC 5332, NGC 2835, NGC 3621, NGC 4298, NGC 4826, NGC 5068, NGC 6744).  
Analysis of the remaining six galaxies (with seven parallel fields: IC 1954, NGC 0685, NGC 1097, NGC 2903-N/S, NGC 5068\footnote{One of the targets re-observed due to guiding failure.  In the first attempt, WFC3 prime observations were corrupted, but most parallel observations were still usuable.}, NGC 7496) yields one additional TRGB constraint.  9.61$\pm$0.39 Mpc is found for NGC 2903-N, and represents the best-available distance for that galaxy.  All PHANGS-HST TRGB measurements and accompanying color-magnitude diagrams will be available from the Extragalactic Distance Database \citep[EDD;][]{EDD,edd21}\footnote{\url{https://edd.ifa.hawaii.edu/}}.  In total, the PHANGS-HST ACS parallel observations have provided eight new TRGB measurements which are the current best-available distances.  The eight measurements span from 4.41$\pm$0.19 Mpc (NGC 4826) to 14.9$\pm$1.4 Mpc (NGC 4298), and are listed in Table~\ref{tab:galaxysample} together with other adopted distances from the literature compiled by \citet{anand20}.


\section{Data Processing Pipeline}
\label{sec:pipeline}

To enable the joint HST-ALMA-MUSE study of star formation in basic units of star clusters, associations, molecular clouds, and HII regions, we have developed an extensive HST data processing pipeline which produces inventories of stars (point sources), compact star clusters, and stellar associations across multiple physical scales in each galaxy.  The pipeline yields aligned, mosaicked images in all five filters as well as catalogs of observed (e.g. photometry and morphological parameters) and physical properties (stellar masses, ages, reddenings, derived through SED fitting).  Ultimately, the PHANGS-HST star cluster and association catalogs will be cross-correlated with the PHANGS-ALMA molecular cloud catalogs \citep[][A.~Hughes et al. in preparation]{phangs-cprops} and PHANGS-MUSE {\sc Hii} region catalogs (F.~Santoro et al. in preparation).  Python packages and high level science products from the HST pipeline are being publicly released and are described in Section~\ref{sec:dataproducts}.

Here, we summarize the overall strategy of the pipeline and provide a framework for subsequent papers which document each of the major components in detail.  The key steps in the PHANGS-HST pipeline are:
\begin{enumerate}
    \item Image drizzling, mosaicking, astrometric calibration (this paper, Section~\ref{sec:drizzle})
    \item Source detection and aperture photometry \citep[][Section~\ref{sec:dolphot}]{thilker21}
    \item Identification of bright, isolated star clusters for determining aperture corrections \cite[][Section~\ref{sec:sinan}]{Deger21}
    \item Derivation of aperture corrections for star clusters \cite[][Section~\ref{sec:sinan}]{Deger21}
    \item Selection of candidate star clusters \citep[][Section~\ref{sec:dave}]{thilker21}
    \item Morphological classification of candidate star clusters \cite[][Section~\ref{sec:brad}]{wei20, whitmore21}
    \item SED fitting \cite[][Section~\ref{sec:jordan}]{turner21}
    \item Identification and photometry of stellar associations (K.~Larson et al. in preparation, Section~\ref{sec:kirsten})
\end{enumerate}

  Our processing workflow reflects common practices for the production of catalogs of compact star clusters in nearby galaxies \citep[e.g.,][]{whitmore10,adamo17}, with the following key augmentations.  First, selection criteria are based on measurement of a series of concentration indices (CI; the difference in photometry measured with circular apertures of two different radii) rather than a single concentration index (Section~\ref{sec:dave}). 
  Second, we inject model star clusters into the HST imaging to aid the definition of selection criteria to separate candidate clusters from point sources and other interlopers; this also provides a foundation to estimate completeness in future work (Section~\ref{sec:dave}).   Third, we utilize convolutional neural network models, as discussed in \cite{wei20}, to supplement human visual inspection, with the goal of eventually automating morphological classification of candidate star clusters, as this has been a limiting step in past cluster studies \citep[e.g.][]{adamo17, whitmore21}.  Fourth, we bifurcate the process for selecting multi-peaked stellar associations from single-peaked compact stellar clusters, by applying a watershed algorithm to identify associations at physical scales from 8~pc to 64~pc (Section~\ref{sec:kirsten}).  This produces a far more complete inventory of young stellar populations, which is crucial for robust comparisons with molecular clouds.
  
\begin{figure*}
\includegraphics[width=\textwidth]{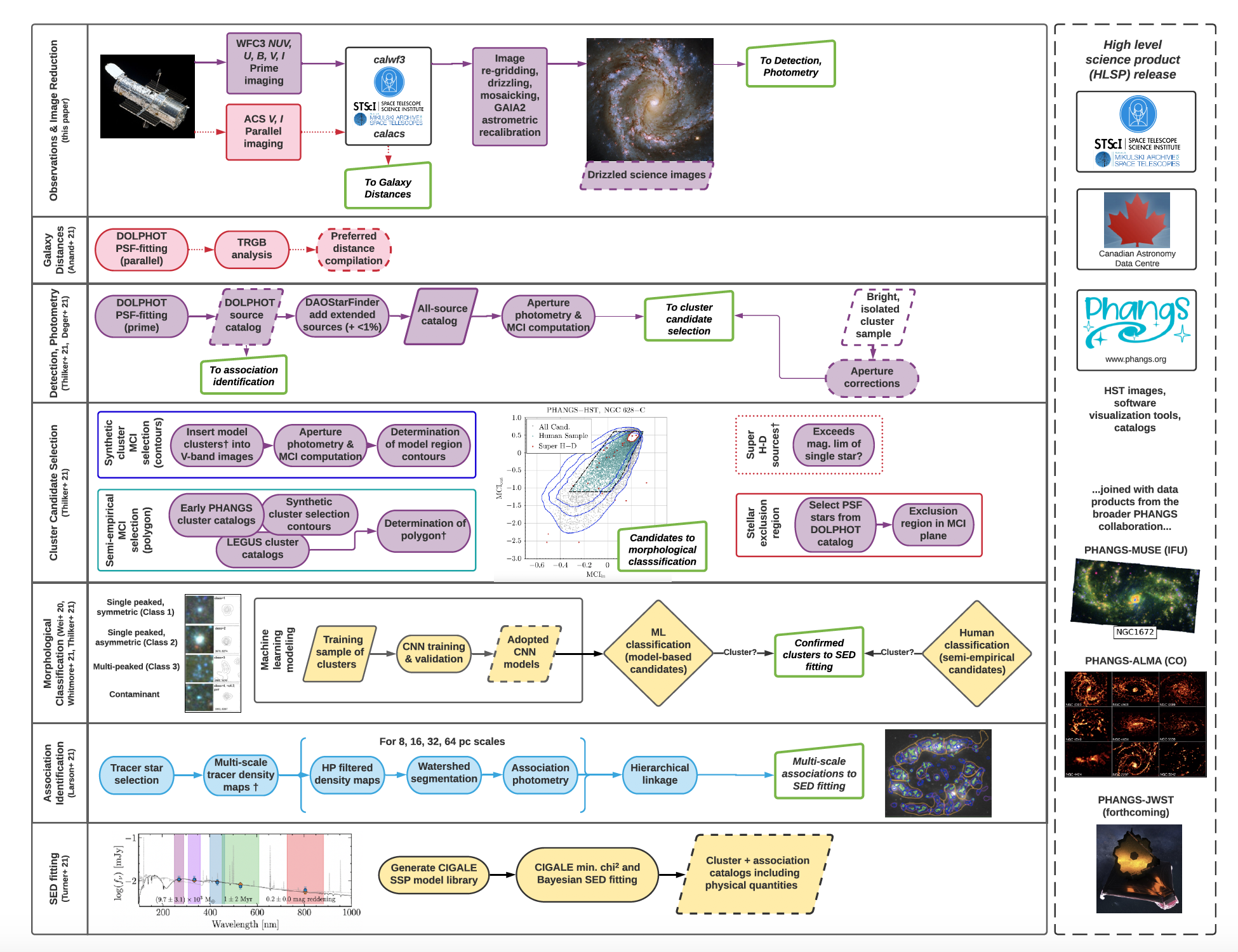}
\caption{Flowchart illustrating the PHANGS-HST star cluster and multi-scale stellar association catalog pipeline.  Chart begins at the upper-left corner with the acquisition of HST imaging, proceeds from left to right along the rows, which represent the major components of the pipeline, and concludes with the production of catalogs at the bottom-right of the chart.  Dashed lines indicate high level science products for public release via MAST, CADC, and the PHANGS collaboration website. $\dagger$indicates that galaxy distance is used in this step.}
\label{fig:flowchart}
\end{figure*}

  A flowchart illustrating the steps in the pipeline as summarized below is provided in Figure~\ref{fig:flowchart}.

\subsection{Drizzling, Mosaicking, Astrometric Calibration}
\label{sec:drizzle}
The process used to drizzle and mosaic the HST imaging data follows current standard procedures.  

Data acquired for PHANGS-HST are first obtained from MAST, along with other suitable archival data taken by previous programs.
These ``FLT'' exposures have been processed through the standard Pyraf/\linebreak[0]{}STSDAS CALACS or CALWFC3 software in the archive, which performs initial data quality flagging, bias subtraction, gain correction, bias stripe removal, correction for CTE losses, dark current subtraction, flat-fielding and photometric calibration, resulting in ``FLC'' FITS files for each ACS/WFC and WFC3/UVIS exposure. 

The PHANGS-HST pipeline is based on the STScI-supported software package DRIZZLEPAC to combine exposures and improve sampling of the PSF, and is used for the prime observations targeting the galaxy disk in a two step drizzle procedure.  The parallel observations are treated separately as discussed in \cite{anand20}. 

The pipeline takes the FLC FITS files retrieved from MAST as input to produce combined images for each filter, which are all aligned and drizzled onto a common grid with pixel scale of $0\farcs04$ (the native WFC3 pixel scale), with astrometry calibrated with GAIA DR2 sources \citep{gaia2, gaia2astrometry}.  The latter is essential for proper alignment of the HST imaging with ALMA and VLT/MUSE data and to facilitate joint study of the three datasets.

The V-band imaging (WFC3 F555W) is used as the reference for positioning of the images in all other filters (NUV, U, B, I) and to define the common pixel grid.  Using the F555W FLC files from MAST for each pointing, the sky positions of the centers and corners of the images are calculated to define a search area to query the ESA DR2 GAIA catalogue.  The TWEAKREG routine then matches the GAIA sources to the objects detected in the F555W drizzled image and calculates average shifts (with accuracy typically better than 0.1 pixel) to correct the astrometric solution.  The number of GAIA sources found in a given F555W HST pointing varies from as few as 15 sources 
to a maximum of 317 
with an average of 40 sources.   The TWEAKBACK routine is then used to propagate the corrected WCS solution back to the original F555W FLC images.  Finally, TWEAKREG and TWEAKBACK are used again, but now with the drizzled images for the other four filters to find sources in common with those detected in F555W, and to align to the F555W image. 

The final drizzle combination is carried out with sky subtraction using the ``globalmin+match'' method for sky calculation. ASTRODRIZZLE first finds a minimum ``global'' sky value for each chip/image extension in all input images and then uses the ``match'' method to compute differences in sky values between images in common sky regions and to equalize the sky values between images.  The final DRC FITS files are in units of e$^{-}$~s$^{-1}$ and are registered with North up and East left as usual.  ``EXP'' (exposure time) and ``ERR'' (error) weight (WHT) maps, calculated from the contribution of all input exposures to a given output pixel, are also produced.  The ERR maps includes all noise sources from detector and sky and are used to compute uncertainties in photometry downstream in the pipeline.

Color composites of the drizzled, mosaicked HST imaging are shown in Figure~\ref{fig:footprints3} for the same six galaxies featured in Figures~\ref{fig:footprints} and~\ref{fig:footprints2}.

\subsection{Source Detection and Photometry}
\label{sec:dolphot}
The principal method of source detection in the PHANGS-HST pipeline is provided by the {\tt DOLPHOT} photometry package \citep[v2.0][]{dolphin02}, which is based on PSF-fitting and operates on the FLC files to detect and deblend sources in HST imaging.  We elect to use {\tt DOLPHOT} for source detection in order to provide a common starting point for the identification of both single-peaked compact star clusters, and multi-peaked stellar associations over a range of physical scales.  The drizzled V-band image is used as the positional reference and sources are detected to 3.5$\sigma$ with PSF-fitting performed at the same image positions in all bands. 

Compact star clusters have effective radii between $0.5$~pc to about $10$~pc \citep{pz10,ryon17}.  At the distances of the galaxies in the PHANGS-HST sample, such objects will have angular sizes close to the HST WFC3 resolution (i.e., 2~pixels $\sim$ 0\farcs08; from 1.7 to 9 pc for the range of distances for the galaxy sample) and will appear sufficiently point-like to be captured by {\tt DOLPHOT}.  To ensure that the catalogs include star clusters with light profiles which may be more extended than the sources detected by {\tt DOLPHOT}, source detection is also performed on the V-band DRC image with {\tt DAOStarFinder} (the python implementation of {\tt DAOPHOT} within the {\tt photutils} astropy-affiliated package), with a kernel FWHM of 2.5 pixels and an effective S/N consistent with the detection threshold used for DOLPHOT.  The merged {\tt DOLPHOT} and {\tt DAOStarFinder} catalogs provide the source lists from which compact star cluster candidates are identified, while the branch of the pipeline which identifies multi-peaked stellar associations (Section~\ref{sec:kirsten}) relies only upon the {\tt DOLPHOT} catalogs, as indicated in the flowchart in Figure~\ref{fig:flowchart}.

Following source detection, the star cluster pipeline then uses the positions from the merged DOLPHOT-DAOStarFinder catalog to perform aperture photometry in all five filters.  For all galaxies photometry is measured in circular apertures with a 4 pixel radius, with background determined in an annulus between 7--8 pixels around the aperture.  An aperture correction (Section~\ref{sec:sinan}) is applied which yields the total fluxes for sources ultimately identified as compact star clusters.  Photometry is also measured in a series of circular apertures (with radii from 1--5 pixels) to compute concentration indices which are used to distinguish star cluster candidates from stars and other sources/\linebreak[0]{}artifacts (Section~\ref{sec:dave}).  Catalogs with these observed parameters are produced for each galaxy, which on average contain about $\sim$500,000 sources. The contribution of non-redundant detections from DAOStarFinder to these catalogs is less than $\sim$1\% of objects with V-band aperture photometry $\mathrm{S/N} >10$, the limit subsequently applied to select star cluster candidates as discussed in Section~\ref{sec:dave}.  

\citet{thilker21} discuss in detail the parameter choices for both the {\tt DOLPHOT} and {\tt DAOStarFinder} routines, and the procedures for obtaining aperture photometry and computing errors.

\subsection{Star Cluster Aperture Corrections}
\label{sec:sinan}

Star clusters, in particular those that are young, are generally found in crowded regions.  Direct, accurate measurement of the total flux is often not possible since the outer light profile is frequently contaminated by other sources.  Thus, HST photometry of compact star clusters in galaxies beyond the Local Group is typically measured with a limited aperture that captures $\sim$50\% of the total flux, and then a correction, determined from bright, isolated clusters, is applied \citep[e.g.,][]{chandar10b, adamo17, cook19}.  

\citet{Deger21} present a detailed discussion of procedures used to determine average aperture corrections and uncertainties for each galaxy.  In summary, the V-band images are visually inspected to identify a few dozen well-detected, isolated, compact clusters, and these objects are used to compute an average correction for each field.  Fixed offsets, based on the change in the encircled energy distributions of the WFC3 PSF with wavelength (which has a minimum FWHM of 0\farcs067 in the V band, and increases to $\sim$0\farcs075 in the NUV and I bands)\footnote{\url{https://www.stsci.edu/itt/APT_help/WFC3/c07_ir07.html}}$^{,}$\footnote{\url{https://www.stsci.edu/hst/instrumentation/wfc3/data-analysis/photometric-calibration/uvis-encircled-energy}} are used to calculate the corresponding corrections for photometry in the NUV, U, B, and I bands, since direct measurements from growth curves in those filters for bright sources in the V-band which are very red or blue can be noisy \citep[][]{cook19, Deger21}.  By construction, the resulting V-band corrections are $\sim$0.75 mag (for the adopted photometric aperture radius of 4 pixels, i.e., $0\farcs16$).  The corrections are larger by 0.19, 0.12. 0.03 and 0.12 mag for the NUV, U, B, and I band, respectively.

\subsection{Star Cluster Candidate Selection}
\label{sec:dave}

 The PHANGS-HST pipeline identifies cluster candidates from the source lists described in Section~\ref{sec:dolphot} using photometric and morphological properties measured in the V-band.  In previous work, the concentration index (computed as the difference between photometry measured in circular apertures with radii of 1 and 3 pixels, CI$_{13}$ or simply CI) has been generally used to remove sources likely to be stars from consideration \citep[e.g.][]{whitmore10,adamo17,cook19}.  To determine the threshold to separate stars from cluster candidates, CIs have been typically measured for a few dozen isolated, bright point sources in an image.  These measurements are then compared with the CI distribution for an analogous sample of compact star clusters (identified through visual inspection and also used to derive aperture corrections). 

For PHANGS-HST, \citet{thilker21} build upon this method to develop candidate selection criteria based on multiple concentration indices (MCI), rather than a single concentration index.  We measure fluxes in circular apertures with radii of 1.0, 1.5, 2.0, 2.5, 3.0, 4.0, and 5.0 pixels, and we define two metrics to characterize the light profiles between 1--2.5 pixels (MCI$_\mathrm{in}$) and 2.5--5 pixels (MCI$_\mathrm{out}$), where
\begin{equation}
\mathrm{MCI} \equiv \frac{1}{3} (\mathrm{NCI}_{ab} + \mathrm{NCI}_{bc} + \mathrm{NCI}_{cd})~,
\label{eq:mci}
\end{equation}
where $a,b,c,d$ represent the radii of the apertures, and
\begin{equation}
\mathrm{NCI}_{ij} \equiv 1 - \frac{\mathrm{CI}_{ij}}{\mathrm{CI}_{ij, \mathrm{fiducial}}}
\end{equation}
is a CI normalized by $\mathrm{CI}_{ij, \mathrm{fiducial}}$, defined to be the CI of a relatively compact cluster; i.e., a PSF-convolved Moffat/EFF87 function \citep[][EFF87]{moffat69,elson1987} with a FWHM of 2 pixels, and a power law slope of 3 describing the surface brightness profile of the extended halo.  The choice of normalization is arbitrary, and enables the various CI to be meaningfully averaged.  With these definitions, clusters measured on HST optical images generally have values of $-0.5\lesssim$ MCI$_\mathrm{in} \lesssim0.2$, $-2\lesssim$ MCI$_\mathrm{out} \lesssim0.6$ (see Figures \ref{fig:empiricalregion} and \ref{fig:modelregion}).  MCI$_\mathrm{in}$ is anti-correlated with the standard CI$_{13}$, as would be expected from their definitions; i.e., more compact sources are characterized by larger MCI values (i.e. more concentrated), which is the opposite of the sense of the standard CI$_{13}$.  By construction, the fiducial cluster lies at the origin of the MCI$_\mathrm{in}$--MCI$_\mathrm{out}$ plane.

We define selection regions in two ways on the MCI$_\mathrm{in}$--MCI$_\mathrm{out}$ plane (Fig.~\ref{fig:modelregion}) to generate two types of candidate lists.  (1) The first set of selection regions are contours in the MCI plane, based on the loci of synthetic star clusters inserted into the V-band imaging for each individual target galaxy.  This selection strategy yields up to several thousand candidates per target, and convolutional neural network models are used for their classification.  (2) The second are smaller polygonal regions, the same set of which are used for all galaxies. Definition of these polygons is based on the loci of human inspected and verified clusters, together with the inner synthetic star cluster contours (which encloses the highest concentration of models) for the first few PHANGS-HST galaxies studied.  This second selection produces smaller samples, more suitable for time-intensive, expert human classification; it is intended to yield an average of $\sim$1000 candidates per field, so that inspection over the full set of 38 PHANGS-HST galaxies is manageable. Once defined, it provides a simpler, faster way of selecting cluster candidates for inspection by bypassing the synthetic star cluster analysis required to generate contours for a given galaxy.  More specifically:

\begin{figure*}
\includegraphics[width=\textwidth]{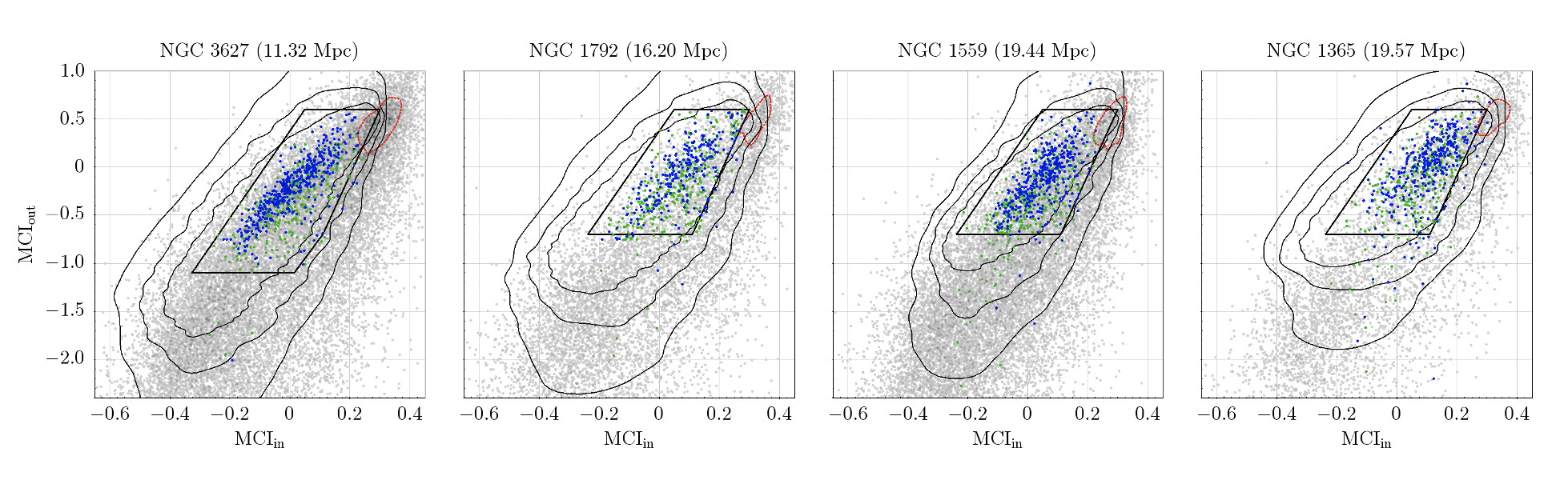}
\caption{Star cluster candidate selection regions defined in the multiple concentration index plane (MCI; see Equation \ref{eq:mci}) based on synthetic star clusters (large contours) for four galaxies at increasing distances from left to right (NGC~3627, NGC~1792, NGC~1559, NGC~1365). Selection regions are also defined semi-empirically (polygon; see also Figure~\ref{fig:empiricalregion}).  Candidates are selected from detections which satisfy basic signal-to-noise criteria (grey), and exclude those in the stellar (point source dominated) region (small red contours).  Candidates within the empirical selection region are slated for visual inspection to a V-band magnitude limit of $\sim24$ mag. The classification of the much larger samples of candidates identified with outermost model contours is automated using convolutional neural network models. In all panels, visually classified class 1 (blue) and class 2 (green) clusters, are shown.   Some clusters appear outside the polygon -- these result from ad-hoc human inspection to confirm that the density of clusters rapidly declines outside this selection region, as well as from the inspection of sources brighter than the Humpherys-Davidson limit.}

\label{fig:modelregion}
\end{figure*}
\begin{figure}
\centering
\includegraphics[width=4in]{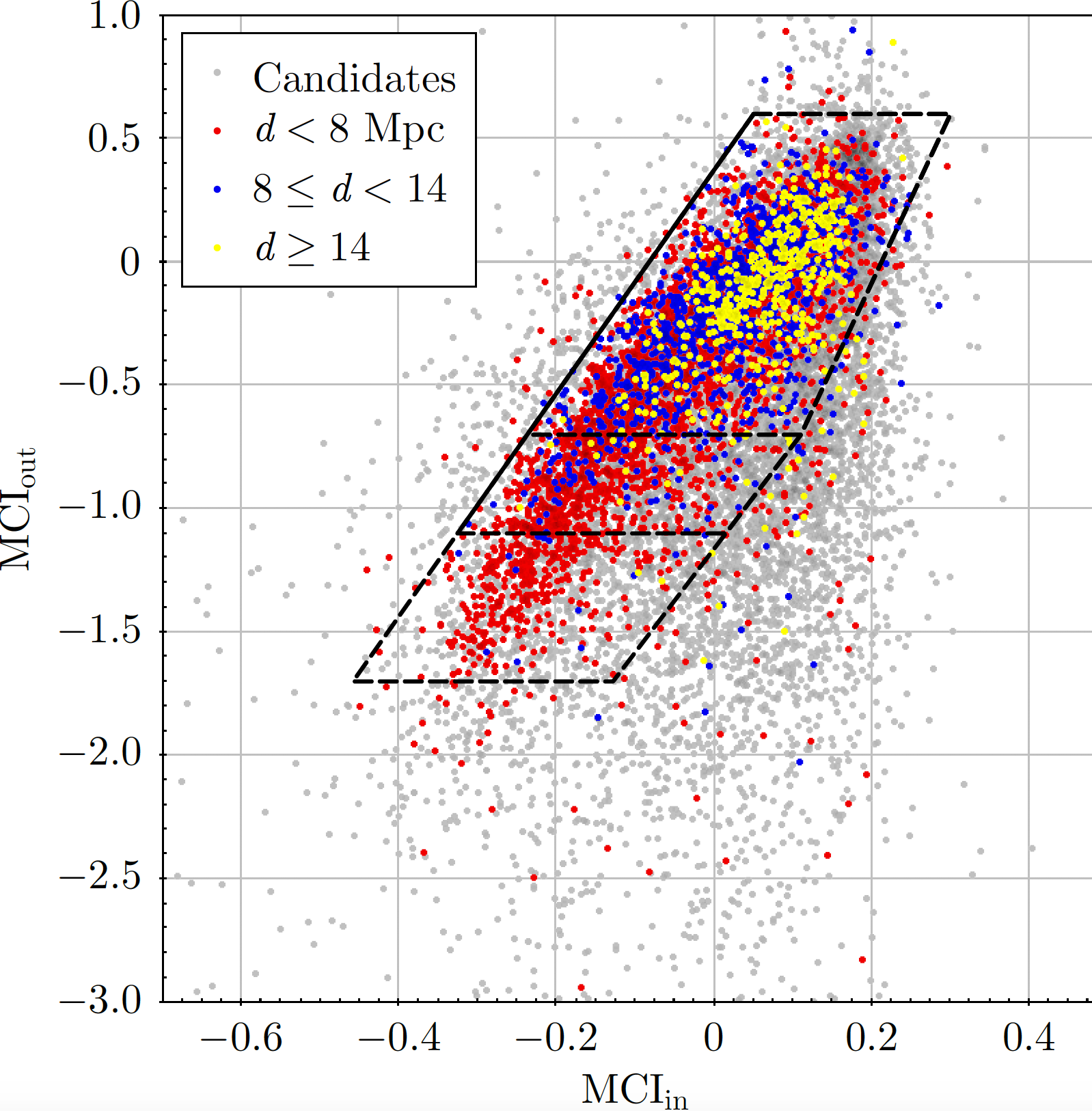}
\caption{Semi-empirically defined selection regions for compact star cluster candidates in the multiple concentration index (MCI; see Equation \ref{eq:mci}) plane.  This selection is devised to produce candidate sample sizes amenable for human inspection, while maximizing the yield of true clusters. We verify that the selection region includes the majority of visually verified class 1 and 2 clusters from 30 galaxies studied by the LEGUS program \citep[][colored points; \url{https://archive.stsci.edu/prepds/legus/dataproducts-public.html}]{calzetti15, adamo17}.  The larger set of candidates from LEGUS are also shown (grey points).  Clusters are color-coded based on the distance of their host galaxy.  The lower bound of MCI$_\mathrm{out}$ increases as the distance of the host galaxy increases (see also Figure~\ref{fig:modelregion}), since the physical resolution decreases and clusters appear more compact.  We adopt different lower bounds for the three different distance ranges indicated in the legend.}
 \label{fig:empiricalregion}
\end{figure}

\begin{enumerate}
\item Selection criteria in the MCI plane are defined based on model star clusters.  Synthetic clusters are generated with Moffat profiles, which again are parameterized by the effective radius (in units of the intrinsic/pre-PSF-convolution FWHM), and the power law slope describing the surface brightness profile of the extended halo.  The modeling includes 216 distinct Moffat profiles, spanning 0.5 pix $\leq \mathrm{FWHM} \leq$ 7 pix and power law slopes from 0.75 to 4. Models are generated for each of these profile types with a distribution of $\sim$4000 apparent magnitudes.  The magnitudes are computed from the distance of the galaxy and V-band luminosities based on solar-metallicity, single-aged stellar population models of \cite{bruzual03}, for a grid of masses ($10^{3} {-} 10^{5}$~\msun), ages (1-1000 Myr), and extinctions ($0 \leq A_\mathrm{V} \leq 0.5$ mag).  The synthetic clusters are randomly inserted into the V-band images 200 at a time, and aperture photometry is performed with the same procedure used to measure real sources.  In total, $\sim4\!\times\!10^6$ clusters are inserted.  The MCI values from these synthetic clusters are plotted on the MCI plane to form a 2D histogram with a bin size of 0.01, and a contour is derived.  Contours are also generated for MCI bin sizes of 0.02 and 0.04, resulting in three nested model selection regions for each field (Fig.~\ref{fig:modelregion}). Typically, the inner model region contains the majority of candidates ($\sim50{-}70\%$), while the two larger contours each add comparable numbers of candidates ($20{-}30$\%).  Currently, the outer contour (i.e., corresponding to the largest bin size) is used to produce candidate lists for automated classification by convolutional neural networks (see Section~\ref{sec:brad}).   

\item Selection criteria in the MCI plane are also defined semi-empirically.  For the first few PHANGS-HST galaxies studied \citep[e.g., NGC~1559,][]{wei20}, sources were inspected over a wide swath of the MCI plane (essentially spanning the outer model contour).  The loci of human verified clusters were used in combination with the tightest model contour to define the left, right, and top edges of the polygon.  Catalogs of visually inspected clusters published by the HST LEGUS program for 34 HST fields in 30 galaxies\footnote{\url{https://archive.stsci.edu/prepds/legus/dataproducts-public.html}} were then used to verify that the majority of LEGUS clusters would be captured by this selection (Figure~\ref{fig:empiricalregion}).  The lower bound of MCI$_\mathrm{out}$ increases as the distance of the host galaxies increases -- clusters will be less resolved and appear more compact at larger distances, and both the model contours and the loci of the human verified cluster populations do indeed shrink with increasing distance.  Using the ensemble of cluster populations identified in the first few PHANGS-HST galaxies studied, plus those in the LEGUS catalogs, we adopt limits of MCI$_{\mathrm{out}} = -1.7, -1.1,-0.7$ for galaxies with distances $\le$ 8 Mpc, 8 Mpc $< d <$ 14 Mpc, $\ge$ 14 Mpc, respectively (Figure~\ref{fig:empiricalregion}). This initial analysis over a wide swath of the MCI plane for PHANGS-HST sources suggests that the density of clusters rapidly drops beyond the boundaries of the polygon, but a more careful investigation of such completeness issues will be the subject of future work.  These semi-empirical polygon selection regions are used to produce candidate lists for human inspection and classification \citep[see Section~\ref{sec:brad} and][]{whitmore21} to a total V-band limit of $\sim$24 mag (corresponding to absolute V-band limits between about $-$5.5 mag and $-$8 mag for the distances of the galaxies in the sample). The exact value of the magnitude limit depends on the number of candidates, since again, the primary purpose of this second selection is to reduce the number of candidates for visual inspection to a manageable level (i.e., $\lesssim$ 1500 per galaxy) while maximizing the yield of true clusters.  So, if the total number of candidates is low (several hundred) then the limit is fainter, whereas if the number of candidates is large, the limit is somewhat brighter.
\end{enumerate}

For both model-based (contours) and semi-empirical (polygon) selection methods:
\begin{itemize}
    \item the regions in the MCI plane dominated by stars (point sources) as detected by {\tt DOLPHOT}, are defined for each field and used to exclude objects.  Figure~\ref{fig:modelregion} shows both the synthetic and semi-empirical cluster selection regions for four galaxies at a range of distances together with their stellar exclusion regions. 
    \item candidates must also satisfy basic criteria: V-band photometry measured in a 4 pixel radius must have $\mathrm{S/N} \geq 10$, and the source must also be detected in at least two other bands with photometric error $\le 0.3$ mag.  The faintest sources in the resulting candidate lists for fields with the standard V-band exposure time of 670~s have total V-band magnitudes of $\sim$24.6, which corresponds to absolute magnitudes between $-4.2$ and $-7.8$ for the distance range spanned by the PHANGS-HST galaxies.
    \item all sources with total, absolute V-band magnitudes brighter than $-10$~mag, and $-0.55 \le MCI_{in} \le$ 0.45 (i.e., $MCI_{in}$ values plausible for real objects) are kept as candidates to help ensure high completeness for the brightest clusters. Sources brighter than $-10$~mag exceed the Humphreys-Davidson (HD) limit, the observed maximum luminosity of individual stars in the LMC, thought to be due to a modified Eddington Limit \citep{hdlimit,lamers17}.  These ``super HD sources" also undergo classification to remove interlopers such as saturated stars and background galaxies.

\end{itemize}
 

\begin{figure}
\centering
\includegraphics[width=4in]{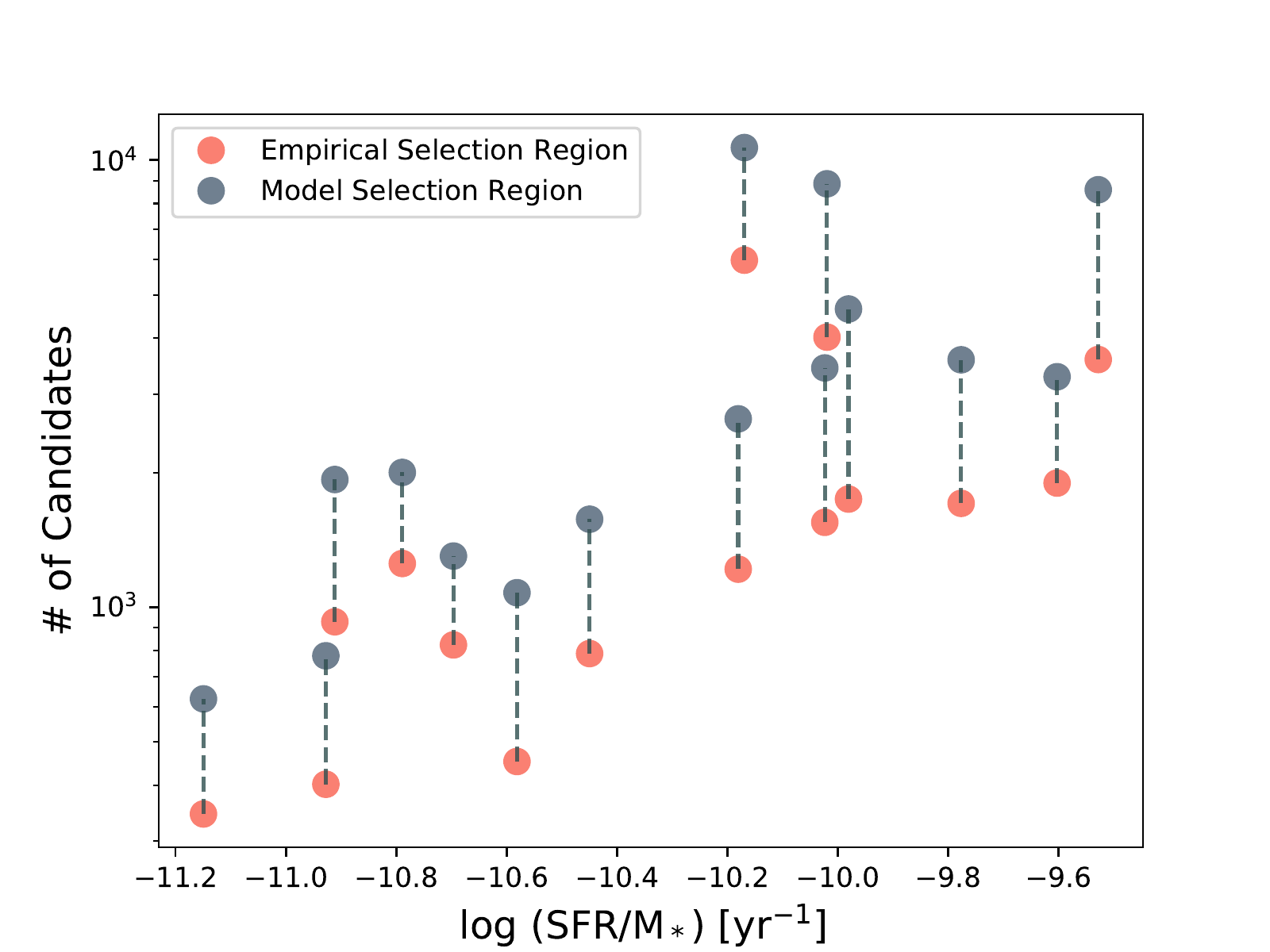}
 \caption{The number of compact star cluster candidates found in the first 15 galaxies processed through the PHANGS-HST pipeline, shown as a function of the sSFR.  Candidates identified using the semi-empirical selection (polygon region) are shown in red, while those resulting from the larger model-based selection contours are shown in grey. The cluster candidates undergo a process of inspection to further remove contaminants and sort the objects into three morphological categories as summarized in Section~\ref{sec:brad}, and described in detail in \cite{wei20}, and \citet{whitmore21}.  The smaller samples resulting from the semi-empirical (polygon) selection are designed for human inspection, while the larger samples from the model contours are fed to convolutional neural networks for automated classification.}
 \label{fig:numcandidates}
\end{figure}

 The number of candidates identified using the empirical MCI selection regions varies from many hundred to several thousand sources for each field, with a median of $\sim$1000 (Figure~\ref{fig:numcandidates}), and should ultimately yield sample sizes of compact star clusters similar to those studied in previous work.  The variation in candidate sample sizes reflects the variation in global sSFRs for the PHANGS-HST galaxies which span a factor of $\sim$10 (Figure~\ref{fig:sample}).  The synthetic cluster MCI selection casts a wider net for potential clusters (Figure~\ref{fig:modelregion}), and yields candidate samples about a factor two larger than the empirical selection (Figure~\ref{fig:numcandidates}).  These larger samples enable analysis of potential incompleteness in previous star cluster studies, in particular for more diffuse clusters which appear to be rare.

These new selection methods, based on the measurement of multiple CI and the use of model star clusters, provides a solid foundation for quantitative investigation of structural properties including: which model clusters actually exist in nature, whether certain clusters are likely to be bound or unbound, and how their morphologies evolve with time.  Our model grid of clusters also facilitates future work to characterize cluster completeness levels \citep[e.g.][]{cliff15}. More discussion on the utility of the approach to cluster selection is provided in \citet{thilker21}. 

\subsection{Star Cluster Candidate Inspection and Morphological Classification}
\label{sec:brad}

The candidate star clusters then undergo a process of inspection to further remove contaminants and sort the clusters into different morphological categories.  A combination of human visual inspection and automated inspection with convolutional neural network models is performed as follows:
\begin{itemize}

\item The smaller lists of candidates resulting from the semi-empirical MCI selection region (polygons) are visually inspected by co-author BCW.  Objects with total V-band magnitude to a limit of $\sim$24 mag receive visual classifications.  

\item The samples of candidates identified in the largest synthetic cluster MCI selection region (contours) are classified by convolutional neural network models, as described below.  
Candidates identified using the semi-empirical selection region generally lie within this synthetic cluster selection area (Figure~\ref{fig:modelregion}), and hence will have both neural network and human visual classifications.

\item Human visual inspection of fainter candidates and those beyond the boundaries of the empirical selection region are performed on an ad-hoc basis to evaluate and monitor the performance of the neural network models.

\end{itemize}


For PHANGS-HST, we adopt the general classification scheme used by LEGUS as described in \citet{adamo17} and \citet{cook19}:
\begin{itemize}
\item Class~1: compact star cluster -- single peak, circularly symmetric, but radial profile more extended relative to point source
\item Class~2: compact star cluster -- similar to Class~1, but elongated or asymmetric 
\item Class~3:compact stellar association -- asymmetric, multiple peaks 
\item Class~4: not a compact star cluster or compact stellar association (e.g. image artifacts, background galaxies, individual stars or pairs of stars)
\end{itemize}

\begin{figure*}
\includegraphics[width=\textwidth]{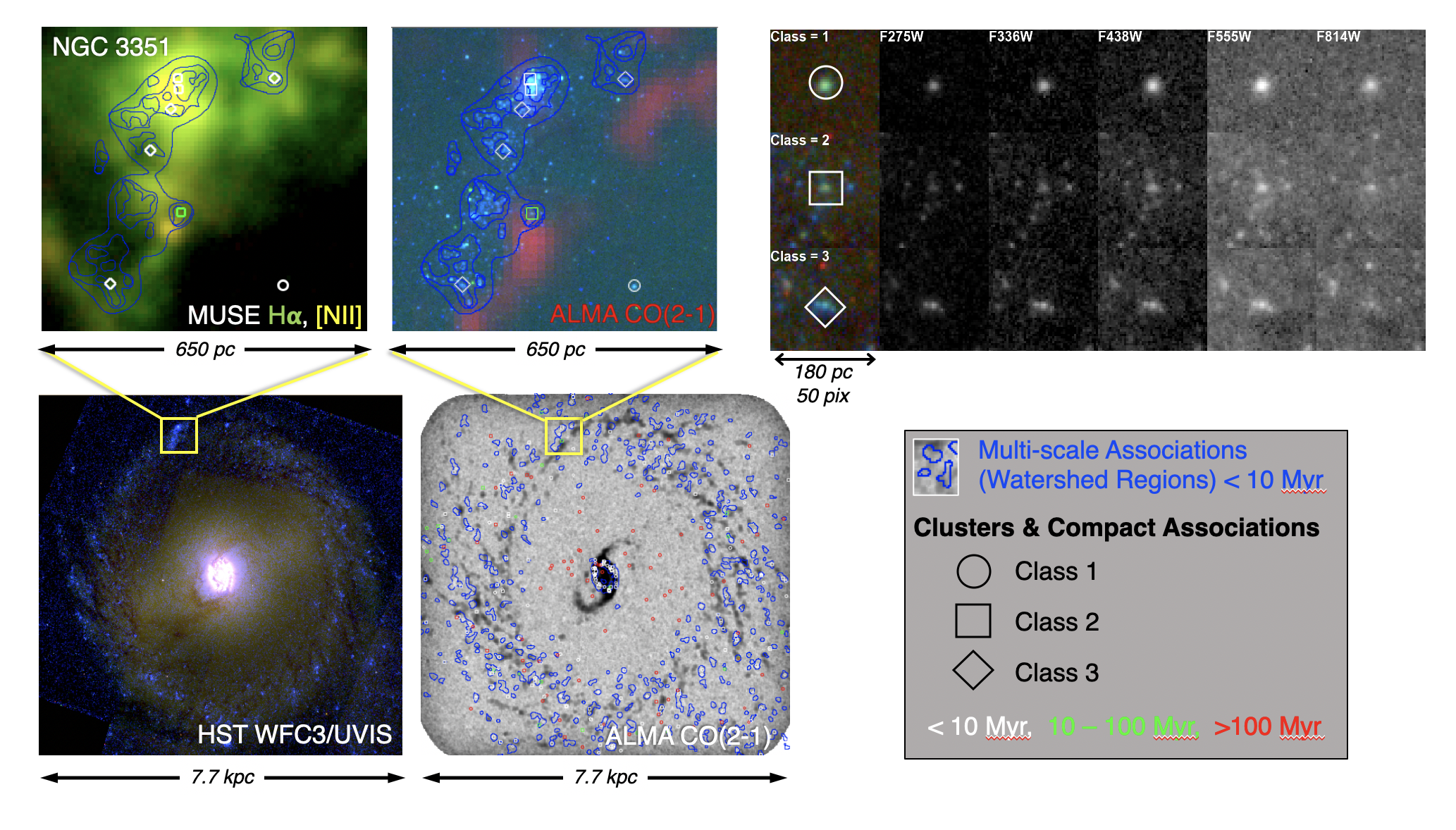}
 \caption{Structures across the physical scales of the star formation hierarchy in NGC 3351, identified by the PHANGS-HST pipeline, from single-peaked compact star clusters, the densest structures, to larger scale multi-peaked stellar associations. \textbf{Bottom left}: Color composite of WFC3/UVIS F275W+F336W (blue), WFC3/UVIS F435W+F555W (green), WFC3/UVIS F814W (red). \textbf{Bottom right}: Young stellar associations ($<$10 Myr) traced by the watershed-based method of K.~Larson et al. (in preparation, blue contours), together with all compact clusters and associations with human visual classifications (Class~1: circles, Class~2: squares, Class~3: diamonds; color coded by age as indicated), overlaid on the PHANGS-ALMA \mbox{CO(2--1)} map.  A 650 pc section of the outer ring (yellow box) is shown in more detail in the top left and middle panels.  All three classes of compact clusters and associations are represented in the selected section, and the magnified view allows all four levels traced by the watershed method (64 pc, 32 pc, 16 pc, 8 pc) to be clearly shown.   \textbf{Top left}: Magnified view using an H$\alpha$ map constructed from the VLT/MUSE IFU data cube.  \textbf{Top middle}: Magnified view using a color composite image where CO is now shown in red.  \textbf{Top right}: Further magnification of 180 pc areas centered on examples of the three classes of compact clusters and associations found in the selected 650 pc section of the outer ring in all PHANGS-HST filters.}
 \label{fig:clusters}
\end{figure*}

Examples of Class 1, 2, 3 objects are shown in Figure~\ref{fig:clusters}.  \citet{whitmore21} provide a detailed description of the process of visual inspection and morphological classification, and discuss differences in the application of this scheme for Class~3 objects relative to the LEGUS project (i.e. we require evidence of 4 or more peaks within a radius of 5 pixels in PHANGS-HST).  A brief history of star cluster classification is also given there \citep[also see][Section 2]{wei20}.
While we continue to include Class~3 objects in our compact cluster catalogs, we note that this is mainly for historical continuity with the LEGUS project.  The PHANGS-HST pipeline is optimized to identify single-peaked compact clusters, and this leads to a high-level of incompleteness for multi-peaked stellar associations.  Instead, we introduce a new identification process for stellar associations, based on a watershed algorithm, which provides a far more complete inventory of young stellar populations and the star formation hierarchy at multiple physical scales, as will be summarized in Section~\ref{sec:kirsten} and presented in K.~Larson et al. (in preparation).

We note that it is debated whether such classifications for a given object 
can distinguish between gravitationally bound clusters and unbound associations, which may form and evolve under distinct conditions \citep{gieleszwart11,kruijssen12,ryon17,krumholz19, ward20}. Statistically however, it is expected that Class~1 should contain the highest percentage of bound clusters, and Class~3 should have the highest percentage of unbound associations \citep[see][for further discussion]{whitmore21}.


Ultimately, PHANGS-HST will generate up to $\sim$80,000 star cluster candidates for inspection and classification.  In previous large studies of star clusters, the process of visual inspection has been a limiting step, which motivated the investigation of automated machine learning techniques \citep[][]{messa18,grasha19,perez20}.  In \cite{wei20}, we studied the application of deep transfer learning techniques to train convolutional neural network (CNN) to classify star cluster candidates according to the scheme above.  Deep transfer learning involves the tuning of a pre-trained network, for example, based on the ImageNet library of everyday objects\footnote{\url{http://www.image-net.org/}}.  In principle, this approach enables CNNs to be successfully trained with relatively small samples (i.e. hundreds to a thousand images), which is the current size of HST star cluster samples with visual classifications.  It provides an alternative to the process of training all network layers from scratch which requires samples which are an order of magnitude larger.  The results of \cite{wei20} were encouraging, as the prediction accuracies (70\%, 40\%, 40--50\%, 50--70\% for Class 1, 2, 3 star clusters, and Class~4 non-clusters, respectively) were found to be competitive with classification consistency between different human classifiers.  The neural network models presented in \cite{wei20} provide a starting point for automated classification of the PHANGS-HST and other HST star cluster candidate samples, which can continue to be optimized.  \citet{whitmore21} present results of the current \cite{wei20} models applied to clusters candidates in five PHANGS-HST galaxies, which also have classifications published by the LEGUS project.  The \citet{whitmore21} analysis includes detailed comparison between human and automated classifications, from overall prediction accuracy to differences in the distribution of ages, UBVI color-color diagrams, and stellar mass functions, and also includes discussion of additional future work to improve performance. 



\subsection{SED Fitting}
\label{sec:jordan}
To derive ages, masses, and reddenings for the sources classified as star clusters and associations, we use a modified version of \textsc{cigale}\footnote{\url{https://cigale.lam.fr}} \citep[\textit{Code Investigating GALaxy Emission};][]{burgarella05,noll09,boquien19}, a publicly available SED fitting package developed for galaxies.  Our modifications, which support the fitting of single-age populations and provide modeling options to facilitate comparison to prior SED modeling results for star clusters, are available in dedicated branches, \texttt{SSP} and \texttt{SSPmag} respectively, of the public \texttt{git} repository\footnote{\url{https://gitlab.lam.fr/cigale/cigale.git}} of \textsc{cigale}.  \cite{turner21} reports on these modifications and the analyses performed to validate the code.  

SED fitting with \textsc{cigale} is performed on the 5-band photometry (NUV-U-B-V-I) for both compact star clusters, and stellar associations.  The fitting is based on the simple (single-aged) population synthesis models of \cite{bruzual03}, assuming solar metallicity and a \cite{chabrier03} IMF (with standard mass limits of $0.1{-}100$~\msun), and no addition of nebular emission.  The \cite{cardelli89} extinction curve with $R_\mathrm{V} = 3.1$ is used and a maximum $\mbox{E(B--V)}=1.5$ mag is imposed.  The reasoning for these choices is discussed in \cite{turner21}.

When CIGALE is run with the same assumptions and theoretical models as used in LEGUS, \citet{turner21} find very similar cluster ages and masses. Quantitatively, fits to identical cluster photometry yield (logarithmic) medians in the ratios of ages and masses for the two surveys of $0.001\pm0.017$~dex and $0.003\pm0.011$~dex, respectively. 

\subsection{Stellar Association Identification}
\label{sec:kirsten} 

The majority of star formation occurs in stellar associations \citep[][and references therein]{lada03,ward18, ward20, wright20}. Compact star clusters, the focus of the previous sections, are formed only in the densest peaks of the star-formation hierarchy \citep{bruce08, kruijssen12} and contain between 1$\sim$50\% of the total star formation in galaxies \citep{kruijssen12, adamo15, cliff16, chandar17, krumholz19, adamo20}.  To produce catalogs of stellar associations, methods distinct from those used to identify single-peaked compact clusters are needed to segment the light distribution over larger physical scales and to probe further into the star-formation hierarchy.  Development of such methods are particularly important for obtaining complete inventories of the youngest stellar populations ($\lesssim$10 Myr), which are requirement for robust joint analysis with molecular clouds and HII regions.

For PHANGS-HST, K.~Larson et al. (in preparation) develop a technique to produce catalogs of stellar associations spanning scales from 8 to 64 pc.  The technique builds upon the watershed routine in the scikit-image python package \citep[\textsc{skimage.segmentation.watershed},][]{vanderwalt14}, which is based on the concept of geological watersheds. The routine identifies regions by ``flooding'' an image, given a set of markers as the starting points.  In addition to the image on which association are to be identified (i.e., requiring segmentation), the inputs needed by \textsc{watershed} are a list of marker positions, and an image mask which defines the areas over which regions are allowed to grow.  

Our adopted technique deploys \textsc{watershed} on a smoothed, filtered map of the positions of point sources (rather than directly on the HST images), and uses a two parameter procedure to determine the marker positions and to produce the image mask from these position maps.
The smoothed, filtered, position maps are produced as follows.  Point sources are selected from the {\tt DOLPHOT} catalogs which satisfy basic requirements on signal-to-noise, sharpness, and data quality.  The positions of these point sources are used to create maps with the same pixel grid as the PHANGS-HST DRC images, containing values of 1's corresponding to the {\tt DOLPHOT} positions, and 0's otherwise.  The maps are then smoothed with Gaussian profiles with FWHM of $2^n$ pc for $n={3,4,5,6}$ (i.e. 8, 16, 32, 64 pc), computed given the distance of the galaxy.  Finally, a high-pass filter is applied by subtracting a map that has been smoothed with a kernel which is four times larger.  

Two parameters are then defined which are tied to the characteristics of a single object on these smoothed, filtered position maps.  The peak threshold parameter is the level above which markers (local maxima) are identified; it is currently defined to be 1.5 times the maximum value for a single object so that the resulting regions are multi-peaked.  The edge threshold parameter is the minimum surface ``brightness'' level beyond which the regions are not allowed to expand further, and used to create a mask image for \textsc{watershed}; it is defined to be the surface ``brightness'' at the FWHM of a single object. 

The smoothed, filtered, position maps enable the identification of structures on the physical scales over which the maps have been smoothed.  The 8 pc smoothed maps allow associations that overlap in size with sources in the compact cluster catalog to be studied, and the maps that are smoothed over larger scales enable the greater star formation hierarchy to be traced.  Structures are identified based on the NUV and V-band {\tt DOLPHOT} point source catalogs.  The resulting NUV-band selected association catalogs will predominantly contain young structures ($\lesssim$100 Myr). 
In comparison, V-band selected catalogs will include structures over a larger range of ages and will facilitate comparison with the compact cluster samples, which have also been V-band selected.  

K.~Larson et al. (in preparation) demonstrate the validity of this technique for identification of stellar associations in the PHANGS-HST galaxy sample based on analysis of NGC 3351 and NGC 1566. 

Figure~\ref{fig:clusters} shows the watershed identified stellar associations in NGC 3351 together with visually classified objects from the compact star cluster pipeline (Section~\ref{sec:dave}).  Three different magnifications of the galaxy are shown to illustrate the full disk and star-forming ring (bottom panels), the multi-scale associations traced by the watershed method (top left and middle panels), and individual compact star clusters and associations (series of postage stamps at top right).  The structures are overlaid on color composites of the HST imaging, ALMA \mbox{CO(2--1)} map, and a color composite of the H$\alpha$ and [NII]$\lambda$6583 maps from MUSE.  Only the youngest ($<$10 Myr) watershed associations are shown to illustrate the correlation with the blue star light, molecular clouds, and {\sc Hii} regions.  Ages are derived using CIGALE as summarized in Section~\ref{sec:jordan} and documented in \cite{turner21}. Details on the procedures used to measure fluxes in the regions are presented in K.~Larson et al. (in preparation).

\begin{figure}
\centering
\includegraphics[width=2.9 in]{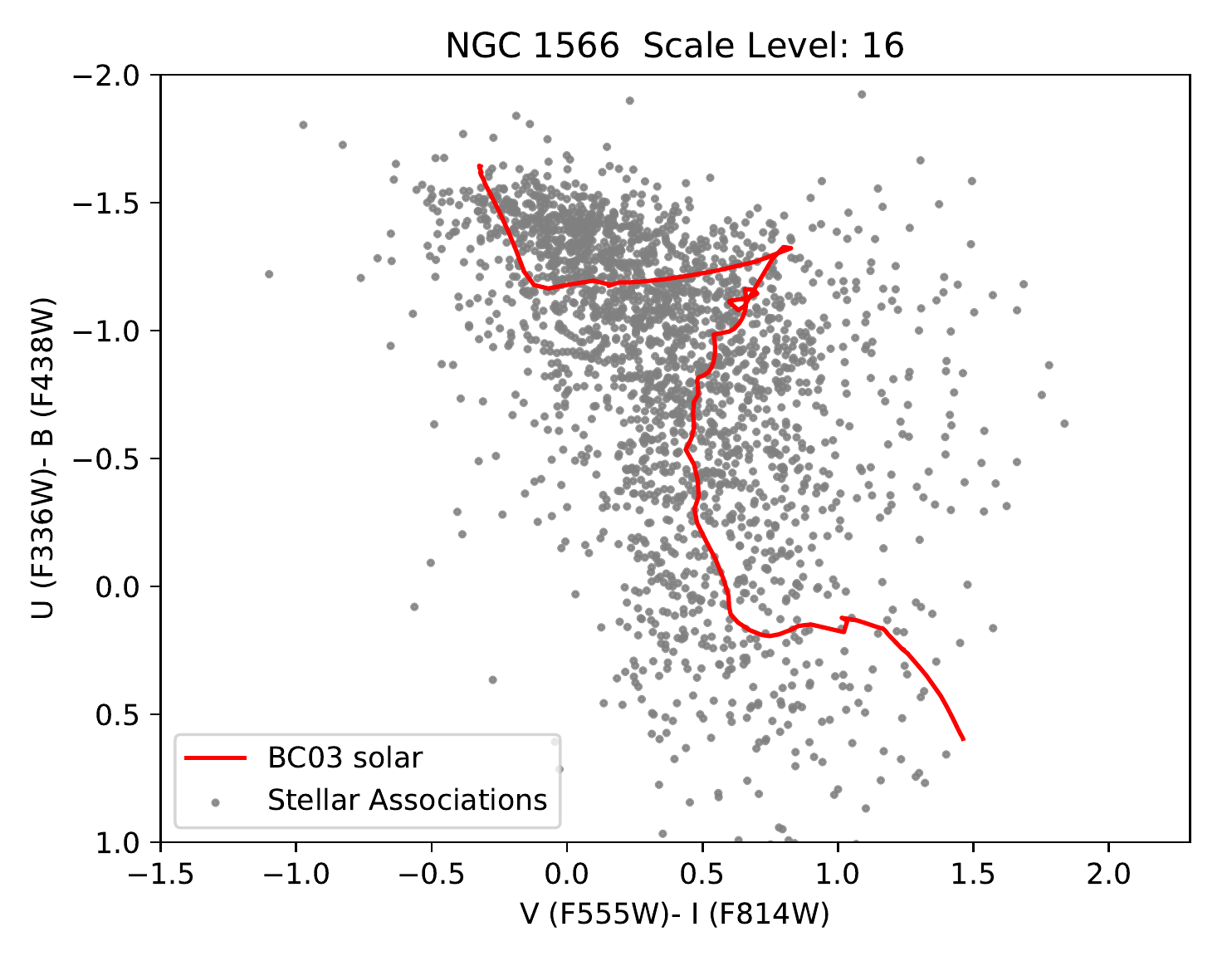}
\includegraphics[width=2.9 in]{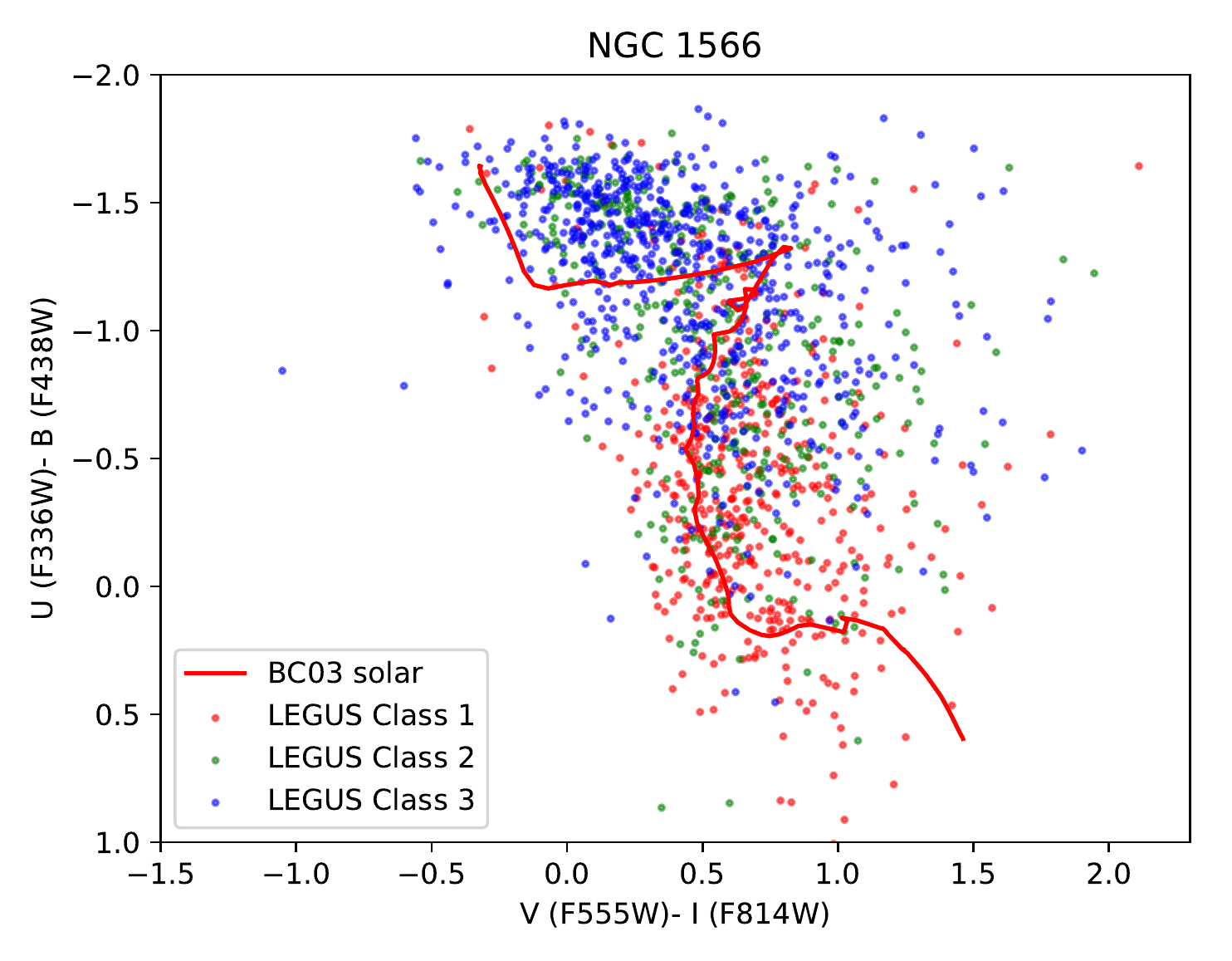}
 \caption{Comparison of UBVI color-color diagrams for multi-peak stellar associations identified with the PHANGS-HST watershed-based procedure (left panel), and compact star clusters and associations from the LEGUS program (right panel).  The stellar associations are based on a V-band map of point source positions smoothed with a 16 pc FWHM gaussian kernel.}
 \label{fig:ngc1566cc}
\end{figure}

\begin{figure}
\includegraphics[width=\columnwidth]{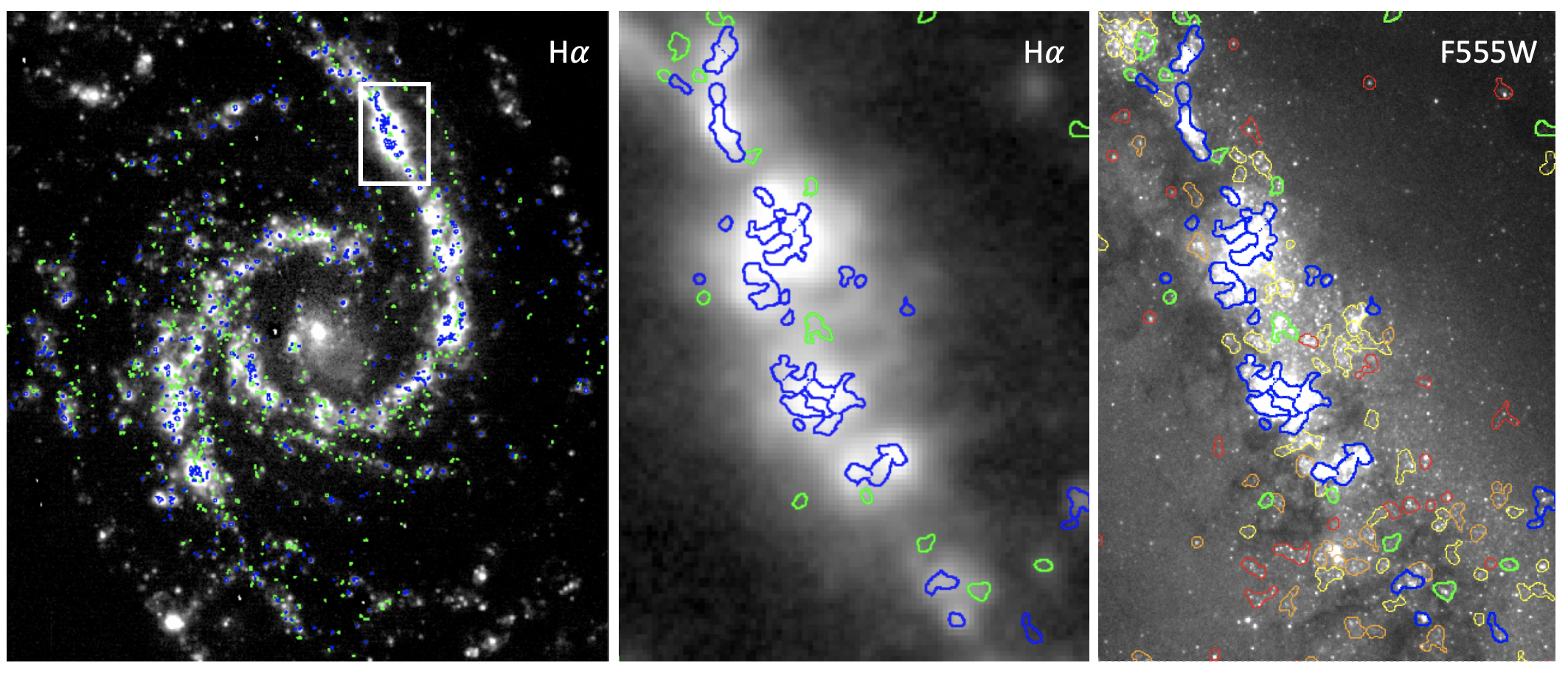}
 \caption{Stellar associations in NGC~1566 color-coded by SED-fit age.  Associations were identified based on a map of F555W point source positions smoothed with a 32~pc FWHM Gaussian kernel, and SED fitting was performed with CIGALE assuming a single-aged stellar population. \textbf{Left:} The youngest associations ($\leq$3~Myr: blue, $3{-}5$~Myr: green) overlaid on PHANGS ground-based H$\alpha$ narrow band imaging (A.~Razza et al. in preparation).  The HST data do not extend to the corners of the imaging, and associations have not been traced in those areas.  \textbf{Middle:} Same as the left panel, but expanded to show greater detail in a portion of the northern spiral arm.  \textbf{Right:}  The same region as the middle panel, but now overlaid on the F555W (V-band) imaging and including older associations ($5{-}10$~Myr: yellow, $10{-}60$~Myr: orange, $>$ 60 Myr red). Age ranges are exclusive of the lower bound and inclusive of the upper bound.}
 
 \label{fig:ngc1566_nuv_v_regions_alma}
\end{figure}

From examination of the properties of the resultant watershed structures, we find:

\begin{itemize}
    \item sample sizes of several hundred up to a few thousand associations in each galaxy.  For associations identified on physical scales comparable to the aperture sizes used for selecting compact clusters candidates (4 pixel radius, which corresponds to 8 pc at a distance of 10 Mpc; Section~\ref{sec:dave}), the numbers of associations identified are about ${\sim}2{-}4$ times larger than the numbers of visually classified compact clusters;  
    \item the process is shown to successfully identify structures at the defined scale. The size distributions are well-defined, and approximately log-normal with medians near the FWHM of the smoothing kernel;
    \item fluxes computed within the boundaries of regions identified on the 8 and 16 pc smoothed images yield colors that are consistent with single-aged stellar population tracks on the UBVI color-color diagram.  The loci are similar to previously studied samples of compact star cluster and associations (Figure~\ref{fig:ngc1566cc}).  The similarity between two panels of Figure 10 demonstrates that these structures, which have been identified with a somewhat complicated algorithm, do behave as one might expect for groups of stars that are physically associated and born close in time.  The associations may not strictly be singled-aged populations, but rather composite populations that can be approximated by a single-aged model. The level of deviation from a single-aged population, particularly as a function of the size of the region, is an issue that will be the topic of future work.  Such work will be facilitated by the PHANGS-HST pipeline’s use of CIGALE, which allows for the self-consistent modeling of both single-aged and composite stellar populations as discussed in \citet{turner21};  
    \item maps of the youngest associations ($<$5 Myr), show excellent correspondence to {\sc Hii} regions observed in narrowband H$\alpha$ imaging.  As the age of the regions increases, they become more anti-correlated with the {\sc Hii} regions, as would be expected (Figure~\ref{fig:ngc1566_nuv_v_regions_alma}).  This provides some evidence that the age-dating of the youngest structures appears to produce reasonable results, despite the simplifying assumption of a single-age population in the SED modeling and the complexities of the overall process to define the regions and measure their photometry.
\end{itemize}


Initial testing of our watershed-based methods for identifying stellar association have been performed with NGC~3351 and NGC~1566.  These two galaxies were chosen as they span a significant range of distances in the sample (from $\sim10-18$ Mpc), and because catalogs of clusters and compact associations are available from LEGUS for comparison.  As with our new methods for identifying clusters, the completeness of our samples of multi-scale associations, particularly possible systematics as a function of distance, will need to be studied with simulations.

\section{Data Products}
\label{sec:dataproducts}
The  PHANGS-HST  dataset  will  enable  science extending well beyond the primary goals of the PHANGS collaboration.
To enable the research community to make full use of the PHANGS-HST data, high-level science products from our star cluster and association catalog pipeline will be released.  The following will be available through the PHANGS homepage at Mikulski Archive for Space Telescopes (MAST)\footnote{\url{https://archive.stsci.edu/hlsp/phangs-hst/}} with digital object identifier \doi{10.17909/t9-r08f-dq31}.  

\subsection{Imaging}

For imaging of the star-forming disk in NUV-U-B-V-I bands for the \ngal\ PHANGS-HST galaxies:
    \begin{itemize}
    \item FLC FITS file for each exposure with astrometric solutions updated based on GAIA DR2 sources.
    \item Combined DRC FITS images of individual pointings in each filter, each drizzled onto a common pixel grid defined for the galaxy target, also with astrometric solutions based on calibration to GAIA DR2 sources.
    \item Mosaicked DRC FITS images in each filter for 13 galaxies covered by multiple pointings (NGC~628, NGC~1097, NGC~1300, NGC~1512, NGC~1672, NGC~2903, NGC~3351, NGC~3621, NGC~3627, NGC~4254, NGC~4321, NGC~4536, NGC~6744).
    \item ERR and EXP weight FITS images for individual pointings as well as mosaics.
\end{itemize}

\subsection{Catalogs}
\begin{itemize}
\item {\tt DOLPHOT} catalogs with 5-band PSF-fitting photometry.
\item Compact star cluster and stellar association candidate catalogs, including position, 5-band aperture photometry, stellar mass, age, reddening, convolutional neural network morphological classification, visual classifications for a subset of candidates in the empricial selection region, multiple concentration indices (MCI) and standard concentration index (CI) values. 
\item Catalogs of stellar associations detected at 8, 16, 32, and 64 pc scales, including 5-band region photometry, stellar mass, age, reddening, effective radius, together with DS9 region files providing peak position and boundaries of regions, and FITS masks of regions.
\end{itemize}

\subsection{Software}
\begin{itemize}
\item The python routines that constitute the PHANGS-HST compact star cluster and association pipeline will be released at \url{https://github.com/PhangsTeam}.  

\item CIGALE augmentations for SED fitting of single-age stellar populations are available in dedicated branches, \texttt{SSP} and \texttt{SSPmag}, respectively, of the public \texttt{git} repository\footnote{\url{https://gitlab.lam.fr/cigale/cigale.git}} of \textsc{cigale}.

\item  Convolutional neural network models for cluster candidate classification as described in \cite{wei20} and \cite{whitmore21}.  An annotated python notebook containing scripts to run the models will be provided. Future updates shown to be improvements over the current models will be released as they are developed . 
\end{itemize}

\subsection{ALMA CO and MUSE Data Products}

PHANGS-ALMA data have been released for the full PHANGS parent sample through the ALMA Archive, the Canadian Astronomy Data Centre (CADC\footnote{\url{https://www.canfar.net/storage/list/phangs/RELEASES/PHANGS-ALMA/}}), and also linked to the PHANGS portal at MAST. 
The PHANGS-ALMA products include the \mbox{$^{12}$CO(2--1)} spectral-line data cubes, signal masks, and derived products such as the integrated intensity, line-of-sight velocity estimate, and spectral line widths.  A description of the data reduction, products, and release is provided in \citet{phangs-alma-pipeline}.

Likewise, PHANGS-MUSE integral field spectrograph data of the 19 galaxies targeted in the course of VLT Large Programme (ESO 1100.B-0651) has been released via the ESO Science Archive\footnote{\url{http://archive.eso.org/scienceportal/home}}, CADC, and also linked to the PHANGS portal at MAST. The released MUSE data includes reduced and fully mosaicked datacubes as well as a series of two-dimensional maps associated with the gas and stellar tracers: broad-band reconstructed images, emission line distribution and kinematics, stellar kinematics, star formation histories (mass and light-weighted age and metallicity maps), extinction maps from Balmer decrement and stellar continuum fitting. The details of the data reduction and analysis processes are provided in E.~Emsellem et al. (submitted).

Links to the archive locations for all released PHANGS products are available at the survey webpage (\url{http://phangs.org/data}).


\section{Summary \& Future Work with JWST}
\label{sec:summary}
For decades, investigations of extragalactic molecular clouds and young resolved stellar populations have proceeded independently, and integrated analysis has been performed only for case studies of select nearby galaxies.  With the transformative capabilities of ALMA and HST working in concert, PHANGS will help bridge the fields of star formation and galaxy evolution by investigating how small-scale physics, which create the basic quanta of star formation, may depend on the physical conditions of the greater galactic environment, and conspire to produce the scaling relationships that characterize the global properties of galaxies.  

With five band NUV-U-B-V-I imaging of the disks of \ngal\ spiral galaxies at distances of 4--23 Mpc, and parallel V and I band imaging of their halos, PHANGS-HST is providing a census of tens of thousands of compact star clusters and associations, which will be combined with PHANGS-ALMA giant molecular clouds (and PHANGS-MUSE {\sc Hii} regions for 19 galaxies in the sample).  Previous to this program, no HST wide-field UV imaging existed for 80\% of the PHANGS-HST sample, and 60\% did not have optical imaging with either WFC3 or ACS.  Thus, PHANGS-HST provides a critical augmentation to the HST archive for nearby spiral galaxies in which both star clusters and molecular clouds can be efficiently detected by HST and ALMA over galactic scales.  Altogether, PHANGS will provide an unprecedented joint catalog of the observed and physical parameters for $\sim$100,000 star cluster, associations, {\sc Hii} regions, and molecular clouds.

In this paper, we have described the ensemble global properties of the \ngal\ galaxy sample targeted for HST observations, and how they were selected from the parent PHANGS-ALMA sample of nearby massive galaxies on the star-forming main sequence.  The acquisition and processing of the HST observations to produce aligned, drizzled, science-ready images were described in detail.  An overview of the major components of the pipeline developed to produce catalogs of single-peak compact star clusters, and a parallel pipeline for multi-scale stellar associations, was provided as a framework for forthcoming detailed papers on each of those components.  We highlight new methods involving multiple concentration index (MCI) parameters, synthetic star clusters, and convolutional neural network models for cluster candidate selection and morphological classifications, as well as a watershed algorithm based procedure for identifying stellar associations from smoothed, filtered maps of point source positions.  We described the data products to be released via MAST at \url{ https://archive.stsci.edu/hlsp/phangs-hst/}, including imaging, catalogs, and software, which will enable community science beyond the main goals of PHANGS.

These data products and the PHANGS census of star clusters, associations, {\sc Hii} regions and molecular clouds will provide the context needed for meaningful study of the earliest phases of dust enshrouded star formation and ISM physics with JWST.  Molecular clouds and UV bright clusters/\linebreak[0]{}associations are the precursors and descendants of the youngest dusty clusters in nearby galaxies to be uncovered by JWST through their infrared emission.   With HST-matched resolution in the near-IR (PSF FWHM $0\farcs066$ at 2~$\mu$m) and order-of-magnitude improved resolution compared to Spitzer in the mid-IR (PSF FWHM of $0\farcs665$ at 21~$\mu$m), molecular clouds with embedded sources can be identified, enabling a key test of our census of inactive clouds, and measurement of time to star formation onset.
 
Through a PHANGS Cycle 1 Treasury program\footnote{\url{https://www.stsci.edu/jwst/phase2-public/2107.pdf}} with an allocation of 106.9 hours, we will obtain NIRCAM and MIRI imaging in eight bands from 2-21~$\mu$m for the 19 galaxies with the full set of PHANGS ALMA, MUSE, and HST observations.  Imaging in the F200W, F300M, and F360M filters will provide a low obscuration view of  stellar photospheric  emission (with some contribution from hot dust to F300M and F360M). F335M, F770W, F1130W will capture PAH emission, tracing a combination of size and charge, with the F300M and F360M filters enabling continuum subtraction for the 3.3~$\mu$m PAH feature. F1000W and F2100W will provide measurements of the warm dust continuum (with some contribution to the F1000W band by silicate absorption).

By resolving the infrared emission across these 19 morphologically diverse galaxies into individual regions and clusters (5--50~pc scales), the PHANGS-JWST observations will complete the inventory of star formation activity in our targets, and reveal the physical state of the small dust grains that heat the ISM. We will calculate mass functions, and spatial distributions for young embedded sources, study their relation to those of other populations, and more clearly identify the conditions that ignite star formation.  The combination of uniform, systematic observations from JWST combined with those already in hand from HST, ALMA, and VLT/MUSE will significantly advance our understanding of the multi-scale process of star formation, and the progression from clouds to visible stars in a galactic context.

\section*{Acknowledgements}

Based on observations made with the NASA/ESA Hubble Space Telescope, obtained from the data archive at the Space Telescope Science Institute. STScI is operated by the Association of Universities for Research in Astronomy, Inc. under NASA contract NAS 5-26555.  Support for Program number 15654 was provided through a grant from the STScI under NASA contract NAS5-26555.

JCL acknowledges the W.M. Keck Institute for Space Studies (KISS) for its support of PHANGS-HST collaboration meetings where key work for this paper was performed.  The PHANGS-HST survey benefited from discussions at the 2014 KISS workshop, ``Bridging the Gap: Observations and Theory of Star Formation Meet on Large and Small Scales.''

JMDK and MC gratefully acknowledge funding from the Deutsche Forschungsgemeinschaft (DFG, German Research Foundation) through an Emmy Noether Research Group (grant number KR4801/1-1) and the DFG Sachbeihilfe (grant number KR4801/2-1) and from the European Research Council (ERC) under the European Union's Horizon 2020 research and innovation programme via the ERC Starting Grant MUSTANG (grant agreement number 714907).

FB and AB acknowledges funding from the European Research Council (ERC) under the European Union’s Horizon 2020 research and innovation programme (grant agreement No.726384/Empire).

RSK and SCOG acknowledge financial support from the DFG via the collaborative research center (SFB 881, Project-ID 138713538) ``The Milky Way System” (subprojects A1, B1, B2, and B8). They also acknowledge subsidies from the Heidelberg Cluster of Excellence {\em STRUCTURES} in the framework of Germany’s Excellence Strategy (grant EXC-2181/1 - 390900948) and funding from the ERC via the ERC Synergy Grant {\em ECOGAL} (grant 855130).

KK, OE, and FS gratefully acknowledge funding from the German Research Foundation (DFG) in the form of an Emmy Noether Research Group (grant No. KR4598/2-1, PI Kreckel).

EW acknowledges support from the DFG via SFB 881 ‘The Milky Way System’ (project-ID 138713538; subproject P2). 

TGW acknowledges funding from the European Research Council (ERC) under the European Union’s Horizon 2020 research and innovation programme (grant agreement No. 694343).

ER acknowledges the support of the Natural Sciences and Engineering Research Council of Canada (NSERC), funding reference number RGPIN-2017-03987.

This paper makes use of the following ALMA data, which have been processed as part of the PHANGS-ALMA survey: \\
ADS/JAO.ALMA\#2012.1.00650.S \linebreak 
ADS/JAO.ALMA\#2013.1.01161.S \linebreak 
ADS/JAO.ALMA\#2015.1.00925.S \linebreak 
ADS/JAO.ALMA\#2015.1.00956.S \linebreak 
ADS/JAO.ALMA\#2017.1.00886.L \linebreak 

This research has made use of the NASA/IPAC Extragalactic Database (NED) which is operated by the Jet Propulsion Laboratory, California Institute of Technology, under contract with NASA. 

\software{PyRAF (Science Software Branch at STScI 2012), Astrodrizzle (Hack et al. 2012), DOLPHOT (v2.0; Dolphin 2002), Photutils (Bradley et al. 2019), CIGALE (Burgarella et al. 2005; Noll et al. 2009; Boquien et al. 2019)}

\bibliographystyle{aasjournal}
\bibliography{all,deep} 

\clearpage
\begin{table*}
\begin{tiny}
\begin{adjustwidth}{-3cm}{}

\begin{tabular}{lllcllllllllll}
\hline
\hline

\textbf{Galaxy} & $\alpha$ & $\delta$ & $b$ &\textbf{D} & \textbf{$\sigma$(D)} & \textbf{Method} & \textbf{D(Ref)}  & $i$ & T & SFR$_{tot}$ & SFR & log M$_{*}$ & $\Sigma_{CO}$\\ 
& [J2000] & [J2000] & [deg] & [Mpc] & [Mpc] & & & [deg]  & & [M$_{\odot}$ yr$^{-1}$] & [M$_{\odot}$ yr$^{-1}$] & [log M$_{\odot}$] & [M$_{\odot}$ kpc$^{-2}$] \\ 
(1) & (2) & (3) & (4) & (5) & (6) & (7) & (8)  & (9) & (10) & (11) & (12) & (13) & (14)\\ 
\hline
IC~1954         & 03h31m31.39s & -51d54m17.4s   & -51.201      & 12.8                 & 2.05             & NAM+TF          & 3+4+5          &	57	&	3.3	&	0.36	& 0.34	&	9.6	&	1.1\\ 
IC~5332$^*$     & 23h34m27.49s & -36d06m03.9s   & -71.366      & 9.01                 & 0.41             & TRGB            & 18             &	27	&	6.8	&	0.41	& 0.11	&	9.5	&	-999\\ 
NGC~628$^*$     & 01h36m41.75s & +15d47m01.2s   & -45.705      & 9.84                 & 0.63             & TRGB            & 2              &	9	&	5.2	&	1.75	& 0.93	&	10.2	&	1.5\\ 
NGC~685         & 01h47m42.81s & -52d45m42.5s   & -62.304      & 19.94                & 2.99             & NAM             & 4+5            &	23	&	5.4	&	0.42	& 0.37	&	9.9	&	0.6\\ 
NGC~1087$^*$    & 02h46m25.16s & -00d29m55.1s   & -51.651      & 15.85                & 2.24             & Group           & 6              &	43	&	5.2	&	1.31	& 1.25	&	9.9	&	1.3 \\ 
NGC~1097        & 02h46m19.05s & -30d16m29.6s   & -64.681      & 13.58                & 2.04             & NAM             & 4+5            &	49	&	3.3	&	4.74	& 3.85	&	10.7	&	2.2 \\ 
NGC~1300$^*$    & 03h19m41.08s & -19d24m40.9s   & -55.223      & 18.99                & 2.85             & NAM             & 4+5            &	32	&	4	&	1.17	& 1.02	&	10.6	&	1.0 \\ 
NGC~1317        & 03h22m44.29s & -37d06m13.3s   & -56.693      & 19.11                & 0.84             & Group           & 2              &	23	&	0.8	&	0.48	& 0.37	&	10.6	&	1.7\\ 
NGC~1365$^*$    & 03h33m36.37s & -36d08m25.4s   & -54.598      & 19.57                & 0.78             & TRGB            & 2              &	55	&	3.2	&	16.90	&12.88 	&	10.8	&	2.7\\ 
NGC~1385$^*$    & 03h37m28.85s & -24d30m01.1s   & -52.706      & 17.22                & 2.58             & NAM             & 4+5            &	44	&	5.9	&	2.09	&1.96	&	9.9	&	1.3\\ 
NGC~1433$^*$    & 03h42m01.55s & -47d13m19.5s   & -51.195      & 15.17                & 1.52             & PNLF            & 7              &	29	&	1.5	&	1.13	&0.56	&	10.4	&	1.3\\ 
NGC~1512$^*$    & 04h03m54.28s & -43d20m55.9s   & -48.166      & 15.94                & 1.59             & PNLF            & 7              &	43	&	1.2	&	1.28	&0.59	&	10.6	&	1.1   \\
NGC~1559        & 04h17m35.77s & -62d47m01.2s   & -41.198      & 19.44                & 0.44             & Mira            & 8              &	65	&	5.9	&	3.76	&3.76	&	10.2	&	1.5\\ 
NGC~1566$^*$    & 04h20m00.42s & -54d56m16.1s   & -43.393      & 17.69                & 2.00             & Group           & 6              &	30	&	4	&	4.54	&3.21	&	10.7	&	2.0   \\ 
NGC~1672$^*$    & 04h45m42.50s & -59d14m49.9s   & -38.990      & 19.40                & 2.91             & NAM             & 4+5            &	43	&	3.3	&	7.60	&6.60	&	10.6	&	2.1   \\ 
NGC~1792        & 05h05m14.45s & -37d58m50.7s   & -36.453      & 16.20                & 2.43             & NAM             & 4+5            &	65	&	4	&	3.70	&3.23	&	10.5	&	1.8\\ 
NGC~2775        & 09h10m20.12s & +07d02m16.6s   &  33.988      & 23.15                & 3.47             & NAM             & 4+5            &	41	&	1.6	&	0.87	&0.70	&	11.1	&	1.2\\ 
NGC~2835$^*$    & 09h17m52.91s & -22d21m16.8s   &  18.509      & 12.22                & 0.94             & TRGB            & 18             &	41	&	5	&	1.24	&0.57	&	9.8	&	0.9     \\ 
NGC~2903        & 09h32m10.11s & +21d30m03.0s   &  44.540      & 9.61                 & 0.39             & TRGB            & 19             &	67	&	4	&	3.08	&2.39	&	10.6	&	1.7 \\ 
NGC~3351$^*$    & 10h43m57.70s & +11d42m13.7s   &  56.368      & 9.96                 & 0.33             & TRGB            & 2              &	45	&	3.1	&	1.32	&0.87	&	10.3	&	1.4     \\ 
NGC~3621        & 11h18m16.51s & -32d48m50.6s   &  26.099      & 7.06                 & 0.28             & TRGB            & 18             &	66	&	6.9	&	0.99	&0.78	&	10.0	&	1.4 \\ 
NGC~3627$^*$    & 11h20m14.96s & +12d59m29.5s   &  64.418      & 11.32                & 0.48             & TRGB            & 2              &	57	&	3.1	&	3.84	&3.31	&	10.7	&	1.9     \\ 
NGC~4254$^*$    & 12h18m49.60s & +14d24m59.4s   &  75.190      & 13.1                 & 2.8              & SCM             & 14             &	34	&	5.2	&	3.07	&2.77	&	10.3	&	2.2     \\ 
NGC~4298        & 12h21m32.76s & +14d36m22.2s   &  75.673      & 14.92                & 1.37             & TRGB            & 18             &	59	&	5.1	&	0.46	&0.43	&	9.9	&	1.4 \\ 
NGC~4303$^*$    & 12h21m54.90s & +04d28m25.1s   &  66.276      & 16.99                & 3.04             & Group           & 6              &	24	&	4	&	5.33	&4.25	&	10.6	&	2.1 \\ 
NGC~4321$^*$    & 12h22m54.83s & +15d49m18.5s   &  76.898      & 15.21                & 0.49             & Cepheid         & 10             &	39	&	4	&	3.56	&2.43	&	10.7	&	1.6 \\ 
NGC~4535$^*$    & 12h34m20.31s & +08d11m51.9s   &  70.641      & 15.77                & 0.37             & Cepheid         & 10             &	45	&	5	&	2.16	&1.23	&	10.5	&	1.2    \\ 
NGC~4536        & 12h34m27.05s & +02d11m17.3s   &  64.730      & 16.25                & 1.13             & TRGB            & 2               &	66	&	4.3	&	3.45	&3.12	&	10.2	&	1.4\\ 
NGC~4548        & 12h35m26.45s & +14d29m46.8s   &  76.830      & 16.22                & 0.38             & Cepheid         & 10              &	38	&	3.1	&	0.52	&0.34	&	10.7	&	1.0\\ 
NGC~4569        & 12h36m49.79s & +13d09m46.6s   &  75.623      & 15.76                & 2.36             & Group           & 6               &	70	&	2.4	&	1.32	&1.22	&	10.8	&	1.4\\ 
NGC~4571        & 12h36m56.38s & +14d13m02.5s   &  76.654      & 14.9                 & 1.2              & Cepheid         & 16              &	33	&	6.4	&	0.29	&0.26	&	10.0	&	0.9\\ 
NGC~4654        & 12h43m56.58s & +13d07m36.0s   &  75.889      & 21.98                & 1.16             & Group           & 10              &	56	&	5.9	&	3.79	&3.06	&	10.5	&	1.4\\ 
NGC~4689        & 12h47m45.56s & +13d45m46.1s   &  76.607      & 15.0                 & 2.25             & NAM+TF          & 3+4+5           &	39	&	4.7	&	0.4	&0.4	&	10.1	&	0.9\\ 
NGC~4826        & 12h56m43.64s & +21d40m58.7s   &  84.423      & 4.41                 & 0.19             & TRGB            & 18              &	59	&	2.2	&	0.2	&0.17	&	10.2	&	1.5\\ 
NGC~5068$^*$    & 13h18m54.81s & -21d02m20.8s   &  41.376      & 5.20                 & 0.21             & TRGB            & 18              &	36	&	6	&	0.28	&0.20	&	9.4	&	0.8    \\ 
NGC~5248        & 13h37m32.02s & +08d53m06.6s   &  68.751      & 14.87                & 1.34             & Group           & 6               &	47	&	4	&	2.29	&1.66	&	10.3	&	1.8\\ 
NGC~6744        & 19h09m46.10s & -63d51m27.1s   & -26.146      & 9.39                 & 0.43             & TRGB            & 18              &	53	&	4	&	2.41	&0.71	&	10.7	&	1.0\\ 
NGC~7496$^*$    & 23h09m47.29s & -43d25m40.6s   & -63.801      & 18.72                & 2.81             & NAM             & 4+5             &	36	&	3.2	&	2.26	&2.00	&	9.8	&	1.2   \\ 

\\
\hline
\hline
\end{tabular}
\end{adjustwidth}
\caption{PHANGS-HST Galaxy Sample\\ 
\textbf{Col 1}: Galaxy name.  $*$ indicates PHANGS-MUSE integral field spectroscopy available, and will be observed in the PHANGS-JWST Treasury Survey. \\
\textbf{Col 2-3}: Right Ascension and Declination.\\
\textbf{Col 4}: Galactic Latitude.\\
\textbf{Col 5-8}: Galaxy distances, uncertainties, and references as follows: \\
1) \cite{2004AJ....127.2031K}
2) \cite{2009AJ....138..332J} 
3) \cite{2016AJ....152...50T} 
4) \cite{2017ApJ...850..207S} 
5) \cite{2020AJ....159...67K} 
6) \cite{2017ApJ...843...16K} 
7) F. Scheuermann et al., in preparation
8) \cite{2020ApJ...889....5H} 
9) \cite{2003ApJ...594..247L} 
10) \cite{2001ApJ...553...47F} 
11) \cite{2010ApJ...715..833O}
12) \cite{2015MNRAS.448.2312B}
13) \cite{2001ApJ...546..681T}
14) \cite{2006ApJ...645..841N}
15) \cite{2019ApJ...886L..27R}
16) \cite{1994Natur.371..385P}
17) \cite{1996ApJ...465L..83R}
18) \cite{anand20}
19) this paper\\
\textbf{Col 9}: Galaxy inclination.  Following \cite{phangs-alma} and adopted from Lang et al. (2020).\\
\textbf{Col 10}: Morphological T-type.\\
\textbf{Col 11}: Total galaxy star formation rate. Based on GALEX FUV and WISE W4 imaging  with SFR prescription calibrated to match results  from population  synthesis  modeling  of \citep{salim16,salim18} as in \citet{phangs-alma}.  For NGC~685 and NGC~4689, GALEX FUV imaging is not available, and the SFRs are based only on WISE W4 data.\\
{Col 12}: Star formation rate, as computed for column 11, but limited to areas studied by PHANGS-HST, as shown in the footprints illustrated in Figure~\ref{fig:footprints2} and provided for the full galaxy sample at \url{https://archive.stsci.edu/hlsp/phangs-hst}.\\
\textbf{Col 13}: Galaxy stellar mass.  Following \citet{phangs-alma}, based on Spitzer IRAC 3.6 $\mu$m when available, or WISE 3.4 $\mu$m, and mass-to-light ratio prescription of \citep{leroy19} calculated as a function of radius in the galaxy.\\
\textbf{Col 14}: Here we calculate $\Sigma_{\rm mol}$ adopting a fixed $\alpha_{\rm CO}^{2-1} = 6.25$~M$_\odot$~pc$^{-2}$~(K~km~s$^{-1}$)$^{-1}$, appropriate for a Galactic conversion factor and a typical CO~(2-1)/CO~(1-0) line ratio \citep{sun18}. Thus, the $x$-axis indicates mean CO surface brightness in units of mass surface density.
}
\label{tab:galaxysample}
\end{tiny}

\end{table*}

\end{document}